

\message{Preloading the phys format:}


\catcode`\@=11

\chardef\f@ur=4
\chardef\l@tter=11
\chardef\@ther=12
\dimendef\dimen@iv=4
\toksdef\toks@i=1 
\toksdef\toks@ii=2
\newtoks\emptyt@ks 

\def\glet{\global\let}
\def\gz@#1{\global#1\z@}
\def\gm@ne#1{\global#1\m@ne}
\def\g@ne#1{\global\advance#1\@ne}

\def\@height{height}       
\def\@depth{depth}         
\def\@width{width}         

\def\@plus{plus}           
\def\@minus{minus}         

\message{macros for text,}


\def\loop#1\repeat{\def\iter@te{#1\expandafter\iter@te \fi}\iter@te
  \let\iter@te\undefined}

\def\bpargroup{\bp@rgroup\ep@r} 
\def\bgrafgroup{\bp@rgroup\ep@rgroup} 
\def\bp@rgroup{\bgroup \let\par\ep@rgroup \let\endgraf}

\def\ep@r{\ifhmode \unpenalty\unskip \fi \p@r}
\def\ep@rgroup{\ep@r \egroup}

\let\p@r=\endgraf   
\let\par=\ep@r
\let\endgraf=\ep@r

\def\lb{\hfil\break}
\def\endpage{\par \vfil \eject}
\def\superendpage{\par \vfil \supereject}



\def\leftline{\@line\hsize\empty\hss}
\def\rightline{\@line\hsize\hss\empty}
\def\centerline{\@line\hsize\hss\hss}

\let\plainrlap=\rlap   
\let\plainllap=\llap   

\def\rlap{\@line\z@\empty\hss}
\def\llap{\@line\z@\hss\empty}

\def\lftline{\@line\hsize\empty\hfil}
\def\rtline{\@line\hsize\hfil\empty}
\def\ctrline{\@line\hsize\hfil\hfil}

\def\@line#1#2#3{\hbox to#1\bgroup#2\let\n@xt#3%
  \afterassignment\@@line \setbox\z@\hbox}
\def\@@line{\aftergroup\@@@line}
\def\@@@line{\unhbox\z@ \n@xt\egroup}

\def\after@arg#1{\bgroup\aftergroup#1\afterassignment\after@@arg\@eat}
\def\after@@arg{\ifcat\bgroup\noexpand\n@xt\else \n@xt\egroup \fi}
\def\@eat{\let\n@xt= } 
\def\eat#1{}           
\def\@eat@#1{\@eat}    


\let\nl=\space

\def\ctrlines#1#2{\par \bpargroup
  \bgroup \parskip\z@skip \noindent \egroup
  \let\ctr@style#1\let\nl\ctr@lines \hfil \ctr@style{#2}\strut
  \interlinepenalty\@M \par}
\def\ctr@lines{\strut \lb \strut \hfil \ctr@style}


\def\begin@{\ifmmode \expandafter\mathpalette\expandafter\math@ \else
  \expandafter\make@ \fi}
\def\make@#1{\setbox\z@\hbox{#1}\fin@}
\def\math@#1#2{\setbox\z@\hbox{$\m@th#1{#2}$}\fin@}

\def\ph@nt{\let\fin@\finph@nt \begin@}
\let\makeph@nt=\undefined
\let\mathph@nt=\undefined

\newif\ift@ \newif\ifb@
\def\topsmash{\t@true\b@false\sm@sh}
\def\botsmash{\t@false\b@true\sm@sh}
\def\smash{\t@true\b@true\sm@sh}
\def\sm@sh{\let\fin@\finsm@sh \begin@}
\let\makesm@sh=\undefined
\let\mathsm@sh=\undefined
\def\finsm@sh{\ift@\ht\z@\z@\fi \ifb@\dp\z@\z@\fi \box\z@}


\newdimen\boxitsep   \boxitsep=5pt

\def\fboxit#1#2{\hbox{\vrule \@width#1\p@
    \vtop{\vbox{\hrule \@height#1\p@ \vskip\boxitsep
        \hbox{\hskip\boxitsep #2\hskip\boxitsep}}%
      \vskip\boxitsep \hrule \@height#1\p@}\vrule \@width#1\p@}}


\begingroup
  \catcode`\:=\active
  \lccode`\*=`\\ \lowercase{\gdef:{*}}   
  \catcode`\;=\active
  \lccode`\* `\% \lowercase{\gdef;{*}}   
  \catcode`\^^M=\active \glet^^M=\space  
\endgroup


\begingroup
  \catcode`\:=\active
  \outer\gdef\comment{\begingroup
    \catcode`\\\@ther \catcode`\%\@ther \catcode`\^^M\@ther
    \catcode`\{\@ther \catcode`\}\@ther \catcode`\#\@ther
    \wlog{* input between `:comment' and `:endcomment' ignored *}%
    \c@mment}
\endgroup
{\lccode`\:=`\\ \lccode`\;=`\^^M
  \lowercase{\gdef\c@mment#1;{\c@@mment:endcomment*}}}
\def\c@@mment#1#2*#3{\if #1#3%
    \ifx @#2@\def\n@xt{\endgroup\ignorespaces}\else
      \def\n@xt{\c@@mment#2*}\fi \else
    \def\n@xt{\c@mment#3}\fi \n@xt}

\message{date and time,}


\newcount\langu@ge

\let\mainlanguage\relax

\begingroup \catcode`\"=\@ther \gdef\dq{"}
  \catcode`\"=\active
  \gdef"#1{\ifx#1s\ss\else\ifx#1SSS\else
    {\accent\dq 7F #1}\penalty\@M \hskip\z@skip \fi\fi}
  \endgroup

\outer\def\english{\gm@ne\langu@ge
  \global\catcode`\"\@ther \glet\3\undefined}
\outer\def\german{\gz@\langu@ge
  \global\catcode`\"\active \glet\3\ss}

\def\case@language#1{\ifcase\expandafter\langu@ge #1\fi}
\def\case@abbr#1{{\let\nodot\n@dot\case@language{#1}.~}}
\def\n@dot{\expandafter\eat}
\let\nodot=\empty


\def\themonth{\xdef\themonth{\noexpand\case@language
  {\ifcase\month \or Januar\or Februar\or M\noexpand\"arz\or April\or
  Mai\or Juni\or Juli\or August\or September\or Oktober\or November\or
  Dezember\fi
  \noexpand\else
  \ifcase\month \or January\or February\or March\or April\or May\or
  June\or July\or August\or September\or October\or November\or
  December\fi}}\themonth}

\def\thedate{\case@language{\else\themonth\ }\number\day
  \case@language{.\ \themonth \else ,} \number\year}

\def\date{\number\day.\,\number\month.\,\number\year}


\def\PhysTeX{$\Phi\kern-.25em\raise.4ex\hbox{$\Upsilon$}\kern-.225em
  \Sigma$-\TeX}

\message{the time,}



\count255=\time \divide\count255 by 60 \edef\thetime{\the\count255 :}
\multiply\count255 by -60 \advance\count255 by\time
\edef\thetime{\thetime \ifnum10>\count255 0\fi \the\count255 }

\message{spacing, fonts and sizes,}


\newskip\refbetweenskip   \newskip\chskiptamount
\newskip\chskiplamount   \newskip\secskipamount
\newskip\footnotebaselineskip   \newskip\interfootnoteskip

\newdimen\chapstretch   \chapstretch=2.5cm
\newcount\chappenalty   \chappenalty=-800
\newdimen\sectstretch   \sectstretch=2cm
\newcount\sectpenalty   \sectpenalty=-400

\def\chskipt{\chapbreak \vskip\chskiptamount}
\def\chskipl{\nobreak \vskip\chskiplamount}
\def\unchskip{\vskip-\chskiplamount}
\def\secskipt{\sectbreak \vskip\secskipamount}
\def\chapbreak{\par \vskip\z@\@plus\chapstretch \penalty\chappenalty
  \vskip\z@\@plus-\chapstretch}
\def\sectbreak{\par \vskip\z@\@plus\sectstretch \penalty\sectpenalty
  \vskip\z@\@plus-\sectstretch}



\begingroup \lccode`\*=`\r
  \lowercase{\def\n@xt#1*#2@{#1}
    \xdef\font@sel{\expandafter\n@xt\fontname\tenrm*@}}\endgroup

\message{loading \font@sel\space fonts,}

=\font@sel r12 
=\font@sel r9
=\font@sel r8
=\font@sel r6

=\font@sel mi12 \skewchar\twelvei='177 
=\font@sel mi9    \skewchar\ninei='177
=\font@sel mi8   \skewchar\eighti='177
=\font@sel mi6   \skewchar\sixi='177

=\font@sel sy10 scaled \magstep1
  \skewchar\twelvesy='60 
=\font@sel sy9    \skewchar\ninesy='60
=\font@sel sy8   \skewchar\eightsy='60
=\font@sel sy6   \skewchar\sixsy='60




=\font@sel bx12 
=\font@sel bx9
=\font@sel bx8
=\font@sel bx6

=\font@sel tt12 
=\font@sel tt8


=\font@sel sl12 
=\font@sel sl9
=\font@sel sl8

\font\twelveit=\font@sel ti12 
\font\nineit=\font@sel ti9
\font\eightit=\font@sel ti8
=\font@sel ti7



=\font@sel csc10 scaled \magstep1 
=\font@sel csc10






\def\twelvepoint{\twelve@point
  \let\normal@spacing\twelve@spacing \set@spacing}
\def\tenpoint{\ten@point
  \let\normal@spacing\ten@spacing \set@spacing}
\def\eightpoint{\eight@point
  \let\normal@spacing\eight@spacing \set@spacing}

\def\rm{\fam\z@ \@fam}
\def\mit{\fam\@ne}
\def\oldstyle{\mit \@fam}
\def\cal{\fam\tw@}
\def\it{\fam\itfam \@fam}
\def\sl{\fam\slfam \@fam}
\def\bf{\fam\bffam \@fam}
\def\tt{\fam\ttfam \@fam}
\def\caps{\@caps}
\def\@fam{\the\textfont\fam}


\def\twelve@point{\set@fonts twelve ten eight }
\def\ten@point{\set@fonts ten eight six }
\def\eight@point{\set@fonts eight six five }

\def\set@fonts#1 #2 #3 {\textfont\ttfam\csname#1tt\endcsname
    \expandafter\let\expandafter\@caps\csname#1csc\endcsname
  \def\n@xt##1##2{\textfont##1\csname#1##2\endcsname
    \scriptfont##1\csname#2##2\endcsname
    \scriptscriptfont##1\csname#3##2\endcsname}%
  \set@@fonts}
\def\set@@fonts{\n@xt0{rm}\n@xt1i\n@xt2{sy}%
  \textfont3\tenex \scriptfont3\tenex \scriptscriptfont3\tenex
  \n@xt\itfam{it}\n@xt\slfam{sl}\n@xt\bffam{bf}\rm}


\def\singlespace{\chardef\@spacing\z@ \set@spacing}
\def\doublespace{\chardef\@spacing\@ne \set@spacing}
\def\triplespace{\chardef\@spacing\tw@ \set@spacing}
\chardef\@spacing=1   

\def\set@spacing{\expandafter\expandafter\expandafter\set@@spacing
  \expandafter\spacing@names\expandafter\@@\normal@spacing
  \normalbaselines}
\def\set@@spacing#1#2\@@#3+#4*{#1#4\multiply#1\@spacing \advance#1#3%
  \ifx @#2@\let\n@xt\empty \else
    \def\n@xt{\set@@spacing#2\@@}\fi \n@xt}

\def\normalbaselines{\lineskip\normallineskip
  \setbaselineskip\normalbaselineskip
  \lineskiplimit\normallineskiplimit}

\def\setbaselineskip{\afterassignment\set@strut \baselineskip}
\def\set@strut{\setbox\strutbox\spacer\z@\baselineskip}

\def\spacer{\hbox\bgroup \afterassignment\x@spacer \dimen@}
\def\x@spacer{\ifdim\dimen@=\z@\else \hskip\dimen@ \fi
  \afterassignment\y@spacer \dimen@}
\def\y@spacer{\setbox\z@\hbox{$\vcenter{\vskip\dimen@}$}%
  \vrule \@height\ht\z@ \@depth\dp\z@ \@width\z@ \egroup}

\def\spacing@names{
  \normalbaselineskip
  \normallineskip
  \normallineskiplimit
  \footnotebaselineskip
  \interfootnoteskip
  \parskip
  \refbetweenskip
  \abovedisplayskip
  \belowdisplayskip
  \abovedisplayshortskip
  \belowdisplayshortskip
  \chskiptamount
  \chskiplamount
  \secskipamount
  }

\def\twelve@spacing{
  14\p@              +5\p@        *
  \p@                +\z@         *
  \z@                +\z@         *
  14\p@              +\p@         *
  20\p@              +\z@         *
  5\p@\@plus\p@      +-2\p@       *
  \z@                +4\p@        *
  8\p@\@plus2\p@\@minus3\p@  +%
    4\p@\@plus3\p@\@minus5\p@     *
  8\p@\@plus2\p@\@minus3\p@  +%
    4\p@\@plus3\p@\@minus5\p@     *
  \p@\@plus2\p@\@minus\p@    +%
    4\p@\@plus3\p@\@minus2\p@     *
  8\p@\@plus2\p@\@minus3\p@  +%
    \p@\@plus2\p@\@minus2\p@      *
  20\p@\@plus5\p@    +\z@         *
  5.5\p@             +\z@         *
  6\p@\@plus2\p@     +\z@         *
  }

\def\ten@spacing{
  11\p@              +4.5\p@      *
  \p@                +\z@         *
  \z@                +\z@         *
  12\p@              +\p@         *
  16\p@              +\z@         *
  5\p@\@plus\p@      +-2\p@       *
  \z@                +5\p@        *
  8\p@\@plus2\p@\@minus3\p@  +%
    4\p@\@plus3\p@\@minus5\p@     *
  8\p@\@plus2\p@\@minus3\p@  +%
    4\p@\@plus3\p@\@minus5\p@     *
  \p@\@plus2\p@\@minus\p@    +%
    4\p@\@plus3\p@\@minus2\p@     *
  8\p@\@plus2\p@\@minus3\p@  +%
    \p@\@plus2\p@\@minus2\p@      *
  20\p@\@plus5\p@    +\z@         *
  5.5\p@             +\z@         *
  6\p@\@plus2\p@     +\z@         *
  }

\def\eight@spacing{
  9\p@               +3.5\p@      *
  \p@                +\z@         *
  \z@                +\z@         *
  10\p@              +\p@         *
  14\p@              +\z@         *
  5\p@\@plus\p@      +-2\p@       *
  \z@                +5\p@        *
  8\p@\@plus2\p@\@minus3\p@  +%
    4\p@\@plus3\p@\@minus5\p@     *
  8\p@\@plus2\p@\@minus3\p@  +%
    4\p@\@plus3\p@\@minus5\p@     *
  \p@\@plus2\p@\@minus\p@    +%
    4\p@\@plus3\p@\@minus2\p@     *
  8\p@\@plus2\p@\@minus3\p@  +%
    \p@\@plus2\p@\@minus2\p@      *
  20\p@\@plus5\p@    +\z@         *
  5.5\p@             +\z@         *
  6\p@\@plus2\p@     +\z@         *
  }

\twelvepoint


\def\large{\par \bgroup \twelvepoint \after@arg\@size}
\def\medium{\par \bgroup \tenpoint \after@arg\@size}
\def\small{\par \bgroup \eightpoint \after@arg\@size}
\def\@size{\par \egroup}

\def\LARGE#1{{\twelve@point #1}}
\def\MEDIUM#1{{\ten@point #1}}
\def\SMALL#1{{\eight@point #1}}

\message{texts, headings and styles,}


\def\submittextone{Zur Ver\"offentlichung in\else Submitted to}
\def\submittexttwo{ eingereicht\else}
\def\abstracthead{Zusammenfassung\else Abstract}
\def\ackhead{Danksagung\else Acknowledgements}
\def\appendixhead{Anhang\else Appendix}
\def\eqabbr{Gl\else eq}
\def\eqsabbr{Gln\else eqs}
\def\figpref{Abb\else Fig}
\def\fighead{Abbildungen\else Figure captions}
\def\figabbr{Bild\nodot\else Fig}
\def\figsabbr{Bilder\nodot\else Figs}
\def\tabpref{Tab\else Tab}
\def\tabhead{Tabellen\else Table captions}
\let\tababbr=\tabpref
\def\tabsabbr{Tab\else Tabs}
\def\refpref{Lit\else Ref}
\def\refhead{Literaturverzeichnis\else References}
\def\refabbr{???\else Ref}
\def\refsabbr{????\else Refs}
\def\tocpref{Inh\else Toc}
\def\tochead{Inhaltsverzeichnis\else Table of contents}
\def\footpref{Anm\else Foot}
\def\foothead{Anmerkungen\else Footnotes}
\def\prfhead{Beweis\else Proof}


\def\UPPERCASE#1{\edef\n@xt{#1}\uppercase\expandafter{\n@xt}}

\let\headlinestyle=\twelverm
\let\footlinestyle=\twelverm
\let\pagestyle=\twelverm       
\let\titlestyle=\bf            
\let\authorstyle=\caps
\let\addressstyle=\sl
\let\sectstyle=\caps           

\let\headstyle=\UPPERCASE      
\let\captionstyle=\it          
\let\journalstyle=\sl
\let\volumestyle=\bf



\def\namrefindent{2em}


\let\footstyle=\empty          

\let\stmttitlestyle=\bf        
\let\stmtstyle=\sl
\let\prftitlestyle=\caps
\let\prfstyle=\sl


\def\skipuserexit{\setbox\z@\box\@cclv}  
\def\shipuserexit{\unvbox\@cclv}         

\def\chapuserexit{\sectuserexit}         
\def\appuserexit{\chapuserexit}          
\def\sectuserexit{\secsuserexit}         
\let\secsuserexit=\relax                 

\message{page numbers and output,}


\newcount\firstp@ge   \firstp@ge=-10000
\newcount\lastp@ge   \lastp@ge=10000
\outer\def\pagesel#1#2{\global\firstp@ge#1 \global\lastp@ge#2
  \wlog{**************************}\wlog{*}%
  \wlog{* don't use \string\p agesel}%
  \wlog{* this will not be supported in future}%
  \wlog{*}\wlog{**************************}%
  \wlog{(* pages #1-#2 selected for printing, others will be skipped *)}}

\newbox\pageb@x

\outer\def\toppagenum{\glet\page@tbn T%
  \glet\headb@x\pageb@x \glet\footb@x\voidb@x}
\outer\def\botpagenum{\glet\page@tbn B%
  \glet\headb@x\voidb@x \glet\footb@x\pageb@x}
\outer\def\nopagenum{\glet\page@tbn N%
  \glet\headb@x\voidb@x \glet\footb@x\voidb@x}

\outer\def\lefthead{\glet\head@lrac L}
\outer\def\righthead{\glet\head@lrac R}
\outer\def\althead{\glet\head@lrac A}
\outer\def\centhead{\glet\head@lrac C}
\outer\def\leftfoot{\glet\foot@lrac L}
\outer\def\rightfoot{\glet\foot@lrac R}
\outer\def\altfoot{\glet\foot@lrac A}
\outer\def\centfoot{\glet\foot@lrac C}

\newtoks\lheadtext   \newtoks\cheadtext   \newtoks\rheadtext
\newtoks\lfoottext   \newtoks\cfoottext   \newtoks\rfoottext

\headline={\headlinestyle \head@foot\skip@head\head@lrac
  \lheadtext\cheadtext\rheadtext\headb@x}
\footline={\footlinestyle \head@foot\skip@foot\foot@lrac
  \lfoottext\cfoottext\rfoottext\footb@x}
\lheadtext={}   \cheadtext={}   \rheadtext={}
\lfoottext={}   \cfoottext={}   \rfoottext={}

\newbox\page@strut
\setbox\page@strut\hbox{\vrule \@height 15mm\@depth 10mm\@width \z@}

\def\head@foot#1#2#3#4#5#6{\unhcopy\page@strut
  \if#1T\hfil \else
    \if#2C\head@@foot{\the#3}{\copy#6}{\the#5}\else
      \if#2A\ifodd\pageno \let#2R\else \let#2L\fi \fi
      \if#2R\head@@foot{\the#4}{\the#5}{\copy#6}\else
        \head@@foot{\copy#6}{\the#3}{\the#4}\fi \fi \fi}
\def\head@@foot#1#2#3{\plainrlap{#1}\hfil#2\hfil\plainllap{#3}}

\let\startpage=\relax  

\outer\def\pageall{\glet\page@ac A%
  \global\countdef\pageno\z@ \global\pageno\@ne
  \global\countdef\pageno@pref\@ne \gz@\pageno@pref
  \glet\page@pref\empty \glet\page@reset\count@
  \glet\chap@break\chskipt \outer\gdef\startpage{\global\pageno}}
\outer\def\pagechap{\glet\page@ac C%
  \global\countdef\pageno\@ne \gz@\pageno
  \global\countdef\pageno@pref\z@ \gz@\pageno@pref
  \gdef\page@pref{\dash@pref}%
  \gdef\page@reset{\global\pageno\@ne \global\pageno@pref}%
  \glet\chap@break\superendpage
  \outer\gdef\startpage##1.{\global\pageno@pref##1\global\pageno}}


\hsize=15 cm   \hoffset=0 mm
\vsize=22 cm   \voffset=0 mm

\newdimen\hoffset@corr@p   \newdimen\voffset@corr@p
\newdimen\hoffset@corrm@p   \newdimen\voffset@corrm@p
\newdimen\hoffset@corr@l   \newdimen\voffset@corr@l
\newdimen\hoffset@corrm@l   \newdimen\voffset@corrm@l

\outer\def\portrait{\switch@pl P%
  \glet\hoffset@corr\hoffset@corr@p
  \glet\voffset@corr\voffset@corr@p
  \glet\hoffset@corrm\hoffset@corrm@p
  \glet\voffset@corrm\voffset@corrm@p}
\outer\def\landscape{\switch@pl L%
  \glet\hoffset@corr\hoffset@corr@l
  \glet\voffset@corr\voffset@corr@l
  \glet\hoffset@corrm\hoffset@corrm@l
  \glet\voffset@corrm\voffset@corrm@l}
\def\switch@pl#1{\if #1\ori@pl \else \superendpage \glet\ori@pl#1%
  \dimen@\ht\page@strut \advance\dimen@\dp\page@strut
  \advance\vsize\dimen@ \dimen@ii\hsize \global\hsize\vsize
  \advance\dimen@ii-\dimen@ \global\vsize\dimen@ii \fi}
\let\ori@pl=P

\def\m@g{\dimen@\ht\page@strut \advance\dimen@\dp\page@strut
  \advance\vsize\dimen@ \divide\vsize\count@
  \multiply\vsize\mag \advance\vsize-\dimen@
  \divide\hsize\count@ \multiply\hsize\mag
  \divide\dimen\footins\count@ \multiply\dimen\footins\mag
  \mag\count@}

\output={\physoutput}

\def\physoutput{\make@lbl
  \ifnum \pageno<\firstp@ge \skipp@ge \else
  \ifnum \pageno>\lastp@ge \skipp@ge \else \shipp@ge \fi \fi
  \advancepageno \skippagenum F\skipheadline F\skipfootline F%
  \ifnum\outputpenalty>-\@MM \else \dosupereject \fi}

\def\skippagenum{\glet\skip@page}
\def\skipheadline{\glet\skip@head}
\def\skipfootline{\glet\skip@foot}

\def\skipp@ge{{\skipuserexit \setbox\z@\box\topins
  \setbox\z@\box\footins}\deadcycles\z@}
\def\shipp@ge{\setbox\pageb@x\hbox{%
    \if F\skip@page \pagestyle{\page@pref \folio}\fi}%
  \dimen@-.5\hsize \advance\dimen@\hoffset@corrm
  \divide\dimen@\@m \multiply\dimen@\mag
  \advance\hoffset\dimen@ \advance\hoffset\hoffset@corr
  \dimen@\ht\page@strut \advance\dimen@\dp\page@strut
  \advance\dimen@\vsize \dimen@-.5\dimen@
  \advance\dimen@\voffset@corrm
  \divide\dimen@\@m \multiply\dimen@\mag
  \advance\voffset\dimen@ \advance\voffset\voffset@corr
  \shipout\vbox{\makeheadline \vbadness\@M \setbox\z@\pagebody
    \dimen@\dp\z@ \box\z@ \kern-\dimen@ \makefootline}}

\def\pagecontents{\ifvbox\topins\unvbox\topins\fi
  \dimen@\dp\@cclv \shipuserexit 
  \ifvbox\footins 
    \vskip\skip\footins \footnoterule \unvbox\footins\fi
  \ifr@ggedbottom \kern-\dimen@ \vfil \fi}

\def\folio{\ifnum\pageno<\z@ \ifcase\langu@ge \MEDIUM{\uppercase
  \expandafter{\romannumeral-\pageno}}\else \romannumeral-\pageno \fi
  \else \number\pageno \fi}

\def\makeheadline{\line{\the\headline}\nointerlineskip}
\def\makefootline{\nointerlineskip \line{\the\footline}}

\skippagenum=F   \skipheadline=F   \skipfootline=F

\message{title page macros,}


\outer\def\titlepage{\glet\titl@fill\vfil}
\outer\def\notitlepage{\gdef\titl@fill{\vskip20\p@}}

\newbox\t@pleft   \newbox\t@pright
\def\t@pinit{%
  \global\setbox\t@pleft\vbox{\hrule \@height\z@ \@width.26\hsize}%
  \global\setbox\t@pright\copy\t@pleft}
\t@pinit

\def\topleft{\t@p\t@pleft}
\def\topright{\t@p\t@pright}
\def\t@p#1#2{\global\setbox#1\vtop{\unvbox#1\hbox{\strut #2}}}

\outer\def\submit#1{\topleft{\case@language\submittextone}%
  \topleft{{#1}\case@language\submittexttwo}}

\let\pubdate=\topright

\outer\def\title{\vbox{\line{\box\t@pleft \hss \box\t@pright}}%
  \skippagenum T\skipheadline T\skipfootline T%
  \t@pinit \titl@fill \vskip\chskiptamount \@title}
\let\titcon=\relax  
\outer\def\titcon{\errmessage{please use \noexpand\nl in the title
  instead of \noexpand\titcon}\unchskip \@title}
\def\@title#1{\ctrlines\titlestyle{#1}\chskipl}
\def\titl@#1{\edef\n@xt{\noexpand\@title{#1}}\n@xt}

\def\author{\aut@add\authorstyle}
\def\autcon{\and@con \author}
\def\address{\aut@add\addressstyle}
\def\addcon{\and@con \address}
\def\and@con{\titl@fill \ctrline{\case@language{und\else and}}}

\def\aut@add#1{\titl@fill \ctrlines{#1\use@nl}}
\def\use@nl{\let\\\use@@nl}
\def\use@@nl{\errmessage{please use \noexpand\nl in addresses and
  (lists of) authors instead of \string\\}\nl}

\def\abstract{\titl@fill \he@d{\case@language\abstracthead}%
  \after@arg\titl@fill}

\def\ack{\chskipt \he@d{\case@language\ackhead}}

\def\he@d#1{\ctrline{\headstyle{#1}}\chskipl}

\message{chapters, sections and appendices,}


\newtoks\l@names   \l@names={\\\the@label}
\let\the@label=\empty

\def\label{\num@lett\@label}
\def\@label#1{\def@name\l@names#1{\the@label}}
\def\quote{\num@lett\empty}

\newinsert\lbl@ins
\count\lbl@ins=0   \dimen\lbl@ins=\maxdimen   \skip\lbl@ins=0pt
\newcount\lbln@m   \lbln@m=0
\let\lbl@saved\empty

\def\pagelabel{\num@lett\@pagelabel}
\def\@pagelabel#1{\ifx#1\undefined \let#1\empty \fi
  \toks@\expandafter{#1}\expandafter\testcr@ss\the\toks@\cr@ss\@@
  \ifcr@ss\else
    \toks@{\cr@ss\lbl@undef}\def@name\l@names#1{\the\toks@}\fi
  \g@ne\lbln@m \insert\lbl@ins{\vbox{\vskip\the\lbln@m sp}}%
  \count@\lbln@m \do@label\store@label#1}
\begingroup \let\save=\relax  
  \gdef\lbl@undef{\message{unresolved \string\pagelabel, use
    \string\save\space and \string\crossrestore}??}
\endgroup
\def\do@label#1{\expandafter#1\csname\the\count@\endcsname}
\def\store@label#1#2{\expandafter\gdef\expandafter\lbl@saved
  \expandafter{\lbl@saved#1#2}}

\def\make@lbl{\setbox\z@\vbox{\let\MEDIUM\relax
  \unvbox\lbl@ins \loop \setbox\z@\lastbox \ifvbox\z@
    \count@\ht\z@ \do@label\make@label \repeat}}
\def\make@label#1{\def\make@@label##1#1##2##3#1##4\@@{%
    \gdef\lbl@saved{##1##3}%
    \ifx @##4@\errmessage{This can't happen}\else
    \def@name\l@names##2{\page@pref\folio}\fi}%
  \expandafter\make@@label\lbl@saved#1#1#1\@@}

\outer\def\lblrestore{\all@restore\l@names}


\let\sect=\relax  \let\s@ct=\relax  

\def\chapinit{\chap@init{\chap@pref}\glet\sect@@eq\@chap@sect@eq
  \sectinit}
\def\appinit{\chap@init{\char\the\appn@m}\glet\sect@@eq\@chap@eq
  \glet\sect@dot@pref\empty \glet\sect@pref\dot@pref
  \glet\sect\undefined}
\def\sectinit{\xdef\sect@dot@pref{\the\sectn@m.}%
  \xdef\sect@pref{\dot@pref\sect@dot@pref}\glet\sect\s@ct}
\def\chap@init#1{\xdef\the@label{#1}\xdef\dot@pref{\the@label.}%
  \xdef\dash@pref{\the@label--}\glet\chap@@eq\@chap@eq}


\newcount\chapn@m   \chapn@m=0

\outer\def\chappage{\glet\chap@page T}
\outer\def\nochappage{\glet\chap@page F}

\outer\def\arabicchapnum{\glet\chap@ar A\gdef\chap@pref{\the\chapn@m}}
\outer\def\romanchapnum{\glet\chap@ar R%
  \gdef\chap@pref{\uppercase{\romannumeral\chapn@m}}}

\let\chap=\relax  \let\ch@p=\relax  

\outer\def\chapters{\glet\chap@yn Y\glet\chap\ch@p
  \chap@init{0}\glet\sect@@eq\@chap@sect@eq \sectinit}
\outer\def\nochapters{\glet\chap@yn N\glet\chap\undefined
  \glet\dot@pref\empty \glet\dash@pref\empty
  \glet\chap@@eq\@eq \glet\sect@@eq\@sect@eq \sectinit}

\outer\def\ch@p#1{\if T\chap@page \superendpage \else \chap@break \fi
  \g@ne\chapn@m \sect@reset \chapinit \page@reset\chapn@m
  \eq@reset \fig@reset \tab@reset
  \toks@{\dot@pref}\toks@ii{#1}\chapuserexit
  \titl@{\the\toks@\ \the\toks@ii}%
  \ifnum\auto@toc>\m@ne \toks@store{#1}\@toc\dot@pref \fi}



\def\sec@title#1#2#3#4#5#6#{\ifx @#6@\g@ne#1\else\global#1#6\fi
  #2\secskipt \xdef\the@label{#3\the#1}\xdef#4{\the@label.}%
  \read@store{\sec@@title#4#5}}
\def\sec@@title#1#2#3#4{\toks@{\the@label.}\toks@ii\toks@store
  #4\bpargroup #2\varitem{\the\toks@}\interlinepenalty\@M
    \let\nl\lb \the\toks@ii \par\nobreak
  \ifnum#3<\auto@toc \@toc{#1}\fi}

\newcount\sectn@m   \sectn@m=0
\def\sect@reset{\gz@\sectn@m}
\outer\def\s@ct{\sec@title\sectn@m
  {\secs@reset \sectinit \eq@@reset \fig@@reset \tab@@reset}%
  \dot@pref\sect@pref{\sectstyle\z@\sectuserexit}}

\let\secs@reset=\relax


\def\sect@lev{\@ne}         
\def\sect@id{sect}          
\def\secs@id{secs}          

\outer\def\newsect{\begingroup \count@\sect@lev
  \let\@\endcsname \let\or\relax
  \edef\n@xt{\new@sect}\advance\count@\@ne
  \xdef\sect@lev{\the\count@\space}\n@xt
  \glet\sect@id\secs@id \xdef\secs@id{\secs@id s}\endgroup}
\begingroup \let\newcount=\relax
  \gdef\new@sect{\wlog{\noexpand\string\secs@nm\@= subsection
      level \noexpand\sect@lev}\noexpand\newcount\secs@nm\n@m@
    \gdef\secs@nm\@reset@{\noexpand\gz@\secs@nm\n@m@}%
    \outer\gdef\secs@nm\@{\noexpand\sec@title\secs@nm\n@m@
      \secs@nm s\@reset@ \csn@me\sect@id\@pref@ \secs@nm\@pref@
      {\secs@nm\style@{\the\count@}\secs@nm\userexit@}}%
    \glet\secs@nm s\@reset@ \relax
    \gdef\secs@nm\style@{\csn@me\sect@id\style@}%
    \gdef\secs@nm\userexit@{\secs@nm s\userexit@}%
    \glet\secs@nm s\userexit@\relax
    \outer\xdef\csn@me toc\secs@id\@
      {\global\auto@toc\noexpand\the\count@\space}%
    \xdef\noexpand\save@@toc{\save@@toc\or\secs@id}}
\endgroup

\def\n@m@{n@m\@}
\def\@pref@{@pref\@}
\def\@reset@{@reset\@}
\def\style@{style\@}
\def\userexit@{userexit\@}

\def\secs@nm{\csn@me\secs@id}
\def\csn@me{\expandafter\noexpand\csname}


\newcount\appn@m   \appn@m=64

\outer\def\appendix{\if T\chap@page \superendpage \else\chap@break \fi
  \g@ne\appn@m \secs@reset \appinit \page@reset\appn@m
  \eq@reset \fig@reset \tab@reset \futurelet\n@xt \app@ndix}

\def\app@ndix{\ifcat\bgroup\noexpand\n@xt \expandafter\@ppendix \else
  \expandafter\@ppendix\expandafter\unskip \fi}

\def\@ppendix#1{\toks@{\case@language\appendixhead~\dot@pref}
  \toks@ii{#1}\appuserexit
  \titl@{\the\toks@\ \the\toks@ii}%
  \ifnum\auto@toc>\m@ne \toks@store{#1}%
  \@toc{\case@language\appendixhead\ \dot@pref}\fi}

\def\app#1{\chskipt \he@d{\case@language\appendixhead\ #1}}

\message{equations,}


\def\num@lett{\cat@lett \num@@lett}
\def\num@@lett#1#2{\egroup #1{#2}}

\def\num@l@tt{\cat@lett \num@@l@tt}
\def\num@@l@tt#1#{\egroup #1}

\def\cat@lett{\bgroup
  \catcode`\0\l@tter \catcode`\1\l@tter \catcode`\2\l@tter
  \catcode`\3\l@tter \catcode`\4\l@tter \catcode`\5\l@tter
  \catcode`\6\l@tter \catcode`\7\l@tter \catcode`\8\l@tter
  \catcode`\9\l@tter \catcode`\'\l@tter}

\def\quote@all#1{\leavevmode\hbox{\mathcode`\-\dq 707B$#1$}}
\def\use{\num@lett\@use}
\def\@use{\setbox\z@\hbox}


\newtoks\e@names   \e@names={}
\newcount\eqn@m   \eqn@m=0

\outer\def\equall{\glet\eq@acs A\glet\eq@pref\empty
  \glet\def@eq\@eq \glet\eq@reset\relax \glet\eq@@reset\relax}
\outer\def\equchap{\glet\eq@acs C\gdef\eq@pref{\dot@pref}%
  \gdef\def@eq{\chap@@eq}\glet\eq@reset\eqz@ \glet\eq@@reset\relax}
\outer\def\equsect{\glet\eq@acs S\gdef\eq@pref{\sect@pref}%
  \gdef\def@eq{\sect@@eq}\glet\eq@reset\eqz@ \glet\eq@@reset\eqz@}
\def\eqz@{\gz@\eqn@m}

\outer\def\equfull{\glet\eq@fs F\gdef\eq@@fs{\let\test@eq\full@eq}}
\outer\def\equshort{\glet\eq@fs S\glet\eq@@fs\relax}

\def\@eq(#1){#1}
\def\@chap@eq{\noexpand\chap@eq\@eq}
\def\@sect@eq{\noexpand\sect@eq\@eq}
\def\@chap@sect@eq{\noexpand\chap@sect@eq\@eq}

\def\chap@eq{\test@eq\empty\dot@pref}
\def\sect@eq{\test@eq\empty\sect@dot@pref}
\def\chap@sect@eq{\test@eq\sect@eq\dot@pref}
\def\test@eq#1#2#3.{\def\n@xt{#3.}\ifx#2\n@xt \let\n@xt#1\fi \n@xt}
\def\full@eq#1#2{}
\def\short@eq#1#2#3.{#1}

\outer\def\equleft{\glet\eq@lrn L\glet\eqtag\leqno
  \glet\eq@tag\leq@no}
\outer\def\equright{\glet\eq@lrn R\glet\eqtag\eqno
  \glet\eq@tag\eq@no}
\outer\def\equnone{\glet\eq@lrn N\glet\eqtag\n@eqno
  \glet\eq@tag\neq@no}

\begingroup
  \catcode`\$=\active \catcode`\*=3 \lccode`\*=`\$
  \lowercase{\gdef\n@eqno{\catcode`\$\active
                \def$${\egroup **}\setbox\z@\hbox\bgroup}}
\endgroup
\def\eq@no{\llap{$\@lign##$}\tabskip\z@skip}
\def\leq@no{\kern-\displaywidth \rlap{$\@lign##$}\tabskip\displaywidth}
\def\neq@no{\@use{$\@lign##$}\tabskip\z@skip}

\def\displaylines{\afterassignment\display@lines \@eat}
\def\display@lines{\displ@y
  \halign\n@xt\hbox to\displaywidth{$\@lign\hfil\displaystyle##\hfil$}%
    &\span\eq@tag\crcr}

\def\eqalignno{\let\eq@@tag\eq@no \eqalign@tag}
\def\leqalignno{\let\eq@@tag\leq@no \eqalign@tag}
\def\eqaligntag{\let\eq@@tag\eq@tag \eqalign@tag}
\def\eqalign@tag{\afterassignment\eqalign@@tag \@eat}
\def\eqalign@@tag{\displ@y
  \tabskip\centering \halign to\displaywidth\n@xt
    \hfil$\@lign\displaystyle{##}$\tabskip\z@skip
    &$\@lign\displaystyle{{}##}$\hfil\tabskip\centering
    &\span\eq@@tag\crcr}

\def\fulltag#1{{\let\test@eq\full@eq#1}}
\def\shorttag#1{{\let\test@eq\short@eq#1}}

\def\eq{\g@ne\eqn@m \make@eq\empty}
\def\make@eq#1{(\eq@pref\the\eqn@m #1)}
\def\EQ{\eq \num@lett\eq@save}
\def\eq@save#1{\def@name\e@names#1{\expandafter\def@eq\make@eq\empty}}

\def\eqn{\eqtag\eq}
\def\EQN{\eqtag\EQ}

\def\eqadv{\g@ne\eqn@m}
\def\EQADV{\eqadv \num@lett\eq@save}

\newcount\seqn@m   \seqn@m=96

\def\subeqbegin{\global\seqn@m96 \subeq}
\def\SUBEQBEGIN{\global\seqn@m96 \SUBEQ}
\def\subeq{\g@ne\seqn@m \make@eq{\char\seqn@m}}
\def\SUBEQ{\num@lett\@SUBEQ}
\def\@SUBEQ#1{\subeq \def@name\e@names#1{\char\the\seqn@m}}

\def\SUBEQNBEGIN{\eqtag\SUBEQBEGIN}

\def\SUBEQN{\eqtag\SUBEQ}

\def\eqapp{\num@lett\@eqapp}
\def\@eqapp#1#2{(\fulltag#1#2)}
\def\eqnapp{\eqtag\eqapp}

\def\queq{\num@lett\@queq}
\def\@queq#1{\quote@all{\eq@@fs(#1)}}
\def\qeq{\case@abbr\eqabbr\queq}
\def\qeqs{\case@abbr\eqsabbr\queq}

\outer\def\eqrestore{\all@restore\e@names}

\message{storage management,}



\newtoks\toks@store
\newtoks\file@list \file@list={1234567}
\def\@tmp{\jobname.$$}
\let\ext@ft=F

\begingroup
  \let\storebox=\relax \let\refnam=\relax  
  \let\RFfile=\relax \let\RFext=\relax     
  \newhelp\opt@help{The options \string\refnam, \string\RFfile\space
    and \string\RFext\space are incompatible with \string\storebox.
    Your request will be ignored.}
  \global\opt@help=\opt@help 
  \gdef\opt@err{{\errhelp\opt@help \errmessage{Incompatible options}}}
\endgroup

\begingroup
  \let\storebox=\relax \let\storelist=\relax  
  \let\storefile=\relax \let\RFfile=\relax    
  \gdef\case@store{%
    \glet\storebox\undefined
    \if B\store@blf \glet\storebox\empty \glet\case@store\case@box
      \else \glet\box@store\undefined
      \glet\box@out\undefined \glet\box@print\undefined
      \glet\box@save\undefined \glet\box@kill\undefined \fi
    \glet\storelist\undefined
    \if L\store@blf \glet\storelist\empty \glet\case@store\case@list
      \else \glet\list@store\undefined
      \glet\list@out\undefined \glet\list@print\undefined
      \glet\list@save\undefined \glet\list@kill\undefined \fi
    \glet\storefile\undefined
    \if F\store@blf \glet\storefile\empty \glet\case@store\case@file
      \else \store@setup
      \glet\file@out\undefined \glet\file@print\undefined
      \glet\filef@rm@t\undefined \glet\fil@f@rm@t\undefined
      \glet\file@save\undefined \glet\file@kill\undefined \fi
    \glet\case@box\undefined \glet\case@list\undefined
    \glet\case@file\undefined \glet\store@setup\undefined
    \case@store}
  \gdef\store@setup{\ifx \RFfile\undefined \glet\file@store\undefined
    \glet\file@open\undefined \glet\file@close\undefined
    \glet\file@wlog\undefined \glet\file@free\undefined
    \glet\file@copy\undefined \glet\file@read\undefined \fi}
\endgroup

\outer\def\storebox{\if T\ext@ft \opt@err \else
    \if L\RF@lfe \if N\ref@sbn \opt@err
    \else \glet\store@blf B\fi \else \opt@err \fi \fi}
\outer\def\storelist{\glet\store@blf L}
\outer\def\storefile{\glet\store@blf F}

\def\read@store{\bgroup \@read@store}
\def\read@@store{\bgroup \catcode`\@\l@tter \@read@store}
\def\@read@store#1{\def\after@read{\egroup \toks@store\toks@i
    #1\after@read \ignorespaces}%
  \catcode`\^^M\active \afterassignment\after@read \global\toks@i}
\def\afterread#1{\bgroup \def\after@read{\egroup #1\after@read}}
\let\after@read=\relax
\begingroup
  \catcode`\:=\active
  \gdef\write@save#1{\write@store{:restore\@type{#1}}}
\endgroup
\def\write@store#1{\s@ve{#1{\the\toks@store}}}

\def\f@rm@t#1#2{\bgroup \ignorefoot
  \leftskip\z@skip \rightskip\z@skip \f@rmat
  \ifx @#1@\everypar{\b@format}\else
    \varitem\@indent{#1}\b@format \fi #2\e@format}
\let\f@rmat=\nointerlineskip
\def\form@t{\unskip \strut \par \@break \eform@t}
\def\b@format{\glet\e@format\form@t \strut}
\def\eform@t{\egroup \glet\e@format\eform@t}
\let\e@format=\eform@t

\def\@store{\case@store\box@store\list@store\file@store}
\def\@sstore#1#2#3{\par \noindent \bgroup \captionstyle
  \case@abbr#2#3:\enskip \the\toks@store \par \egroup \@store#1{#3.}}
\def\@add#1{\read@store{\@store#1{}}}
\def\@out#1#2#3#4#5{\case@store\box@out\list@out\file@out#1\begingroup
  \if T#2\let\chap@break\superendpage \fi \chap@break
  \chap@init{\case@language#3}%
  \if C\page@ac \skippagenum T\fi
  \page@reset6#1%
  \@style \he@d{\strut\case@language#4}\@break \@print#1%
  \ifx\chap@break\superendpage \superendpage \fi
  \def\\##1{\glet##1\undefined}\the#5\global#5\emptyt@ks
  \endgroup \fi}
\def\@print{\case@store\box@print\list@print\file@print}
\def\@save{\case@store\box@save\list@save\file@save}
\def\@kill{\case@store\box@kill\list@kill\file@kill}
\def\@ext#1#2#3 {\if B\store@blf \opt@err \else
  \glet\ext@ft T\@add#1{#2#3 }\bgroup
   \def\@store##1##2{}\input#3 \egroup \fi}
\def\@@ext#1#2{\let#1#2\everypar\emptyt@ks
  \def\read@store##1{\relax##1}\def\@store##1{\f@rm@t}\input}

\def\case@box#1#2#3{\case@@store#1%
  \fig@box\tab@box\ref@box\toc@box\foot@box}
\def\case@list#1#2#3{\case@@store#2%
  \fig@list\tab@list\ref@list\toc@list\foot@list}
\def\case@file#1#2#3{\case@@store{\expandafter#3}%
  \fig@file\tab@file\ref@file\toc@file\foot@file}

\def\case@@store#1#2#3#4#5#6#7{\ifcase#7%
  \toks@{\fig@type#1#2}\or
  \toks@{\tab@type#1#3}\or
  \toks@{\ref@type#1#4}\or
  \toks@{\toc@type#1#5}\or
  \toks@{\foot@type#1#6}\fi
  \expandafter\let\expandafter\@type\the\toks@}
\def\@style{\csname\@type style\endcsname}
\def\@indent{\csname\@type indent\endcsname}
\def\@break{\csname\@type break\endcsname}

\def\box@store#1#2{\global\setbox#1\vbox
  {\ifvbox#1\unvbox#1\fi \@style \f@rm@t{#2}{\the\toks@store}}}
\def\box@out{\ifvbox}
\def\box@print{\vskip\baselineskip \unvbox}
\begingroup
  \catcode`\:=\active \catcode`\;=\active
  \gdef\box@save#1{\wlog{; Unable to save text for
    \@type's with option :storebox}}
\endgroup
\def\box@kill#1{{\setbox\z@\box#1}}

\def\list@store#1#2{\toks@\expandafter{#1\\}%
  \xdef#1{\the\toks@ {#2}{\the\toks@store}}}
\def\list@out#1{\ifx #1\empty \else}
\def\list@print#1{\let\\\f@rm@t #1\glet#1\empty}
\def\list@save{\def\\##1##2{\toks@store{##2}\write@save{##1}}%
  \newlinechar`\^^M}
\def\list@kill#1{\glet#1\empty}

\begingroup
  \catcode`\:=\active
  \gdef\file@store#1#2#3#4{%
    \if0#3\expandafter\file@open\the\file@list\@@#1#2\fi
    {\newlinechar`\^^M\let\save@write#2\write@store{::{#4}}}}
\endgroup
\def\file@open#1#2\@@#3#4{\immediate\openout#4\@tmp#1
  \gdef#3{#3#4#1}\global\file@list{#2}\file@wlog{open}#1}
\def\file@wlog#1#2{\wlog{#1 \@tmp#2 for \@type's}}
\def\file@out#1#2#3{\if0#3\else}
\def\file@print#1#2#3{\file@close#2#3\let\\\filef@rm@t
  \file@copy#1#2#3}
\def\filef@rm@t#1{\bgroup \catcode`\@\l@tter \fil@f@rm@t{#1}}
\def\fil@f@rm@t#1#2{\egroup \f@rm@t{#1}{#2}}
\def\file@save#1#2#3{\if0#3\else \file@close#2#3%
  \def\\##1{\read@@store{\expandafter\file@store#1{##1}%
    \write@save{##1}}}%
  \newlinechar`\^^M\file@copy#1#2#3\fi}
\def\file@kill#1#2#3{\if0#3\else \file@close#2#3\file@free#1#2#3\fi}
\def\file@close#1#2{\immediate\closeout#1\file@wlog{close}#2}
\def\file@copy#1#2#3{\file@free#1#2#3\file@read#3}
\def\file@read#1{\input\@tmp#1 }
\def\file@free#1#2#3{\gdef#1{#1#20}%
  \global\file@list\expandafter{\the\file@list#3}}

\message{figures,}


\def\if@t#1#2#3#4#5#6{\glet#1#6\gdef#2{\dot@pref}\glet#3#5\glet#4\relax
  \if#6A\glet#2\empty \glet#3\relax \fi
  \if#6S\gdef#2{\sect@pref}\glet#4#5\fi}
\def\bf@t#1{\let\f@t#1\num@l@tt}
\def\ef@t#1#2#3#4#{\ifx @#4@\g@ne#1\xdef\thef@tn@m{#2\the#1}\else
  \gdef\thef@tn@m{#4}\fi #3\read@store\f@t}
\def\@extf@t#1#2#{\gdef\thef@tn@m{#1}\f@t}


\newtoks\f@names   \f@names={}
\newcount\fign@m   \fign@m=0
\def\fig@type{fig}
\newbox\fig@box
\let\fig@list=\empty
\newwrite\fig@write  \def\fig@file{\fig@file\fig@write0}

\outer\def\figall{\fig@init A}
\outer\def\figchap{\fig@init C}
\outer\def\figsect{\fig@init S}
\def\fig@init{\if@t\fig@acs\fig@pref\fig@reset\fig@@reset\figz@}
\def\figz@{\gz@\fign@m}

\outer\def\figpage{\glet\fig@page T}
\outer\def\nofigpage{\glet\fig@page F}

\def\fig{\bf@t\fig@\@fig}
\def\FIG{\bf@t\fig@\@FIG}
\def\ffig{\bf@t\ffig@\@fig}
\def\FFIG{\bf@t\ffig@\@FIG}

\def\@fig{\ef@t\fign@m\fig@pref\relax}
\def\@FIG#1{\ef@t\fign@m\fig@pref{\def@name\f@names#1{\thef@tn@m}}}

\def\fig@{\@store0{\thef@tn@m .}}
\def\ffig@{\@sstore0\figpref{\thef@tn@m}}
\def\figadd{\@add0}

\def\qufig{\case@abbr\figabbr\num@lett\quote@all}
\def\qufigs{\case@abbr\figsabbr\num@lett\quote@all}

\outer\def\figout{\@out0\fig@page\figpref\fighead\emptyt@ks}
\outer\def\figkill{\@kill0}
\outer\def\restorefig#1{\read@store{\@store0{#1}}}
\outer\def\figrestore{\all@restore\f@names}

\outer\def\FIGext{\@ext0\FIG@ext}
\def\FIG@ext{\@@ext\@FIG\@extf@t}


\newdimen\spictskip  \spictskip=2.5pt

\def\pict{\bf@t\pict@\@fig}
\def\PICT{\bf@t\pict@\@FIG}

\def\pict@{\vskip\the\toks@store \bpargroup
  \raggedright \captionstyle \varitem{\qufig{\thef@tn@m\,}:}%
  \let\@spict\spict@ \spacefactor998\ignorespaces}

\def\spict#1{\ifnum\spacefactor=998\else \parvskip\spictskip \fi
  \@spict{#1\enskip}\ignorespaces}
\def\spict@#1{\setbox\z@\hbox{#1}\advance\hangindent\wd\z@
  \box\z@ \let\@spict\llap}


\outer\def\graphics{\glet\graphic\gr@phic}
\outer\def\nographics{\glet\graphic\nogr@phic}

\def\gr@phic#1{\vbox\bgroup \def\gr@@@ph{#1}\bfr@me\gr@ph}
\def\nogr@phic#1{\vbox\bgroup \write\m@ne{Insert plot #1}\bfr@me\fr@me}
\def\frame{\vbox\bgroup \bfr@me\fr@me}
\def\bfr@me#1#2#3#4{\tfr@me\z@\dimen@iv#2\relax 
  \dimen@#3\relax \tfr@me\dimen@\dimen@ii#4\relax 
  \setbox\z@\hbox to\dimen@iv{#1}\ht\z@\dimen@ \dp\z@\dimen@ii
  \box\z@ \egroup}
\def\tfr@me#1#2#3\relax{#2#3\relax \ifdim#2<-#1\errhelp\fr@mehelp
  \errmessage{Invalid box size}#2-#1\fi}
\newhelp\fr@mehelp{The \string\wd\space and \string\ht+\string\dp\space
  of a \string\frame\space or \string\graphic\space \string\box\space
  must not be negative and will be changed to 0pt.}

\def\gr@ph{\lower\dimen@ii\gr@@ph b\hfil
  \raise\dimen@\gr@@ph e}
\def\gr@@ph#1{\hbox{\special{^X\gr@@@ph^A}}}         
\def\gr@@ph#1{\hbox{\special{#1plot GKSM \gr@@@ph}}} 
\def\gr@@ph#1{\hbox{\special{^X#1plot GKSM \gr@@@ph^A}}} 
\def\fr@me{\vrule\fr@@@me
  \bgroup \dimen@ii-\dimen@ \fr@@me\dimen@ii \hfilneg
  \bgroup \dimen@-\dimen@ii \fr@@me\dimen@ \vrule\fr@@@me}
\def\fr@@me#1{\advance#1.4\p@ \leaders \hrule\fr@@@me \hss \egroup}
\def\fr@@@me{\@height\dimen@ \@depth\dimen@ii}

\message{tables,}


\newtoks\t@names   \t@names={}
\newcount\tabn@m   \tabn@m=0
\def\tab@type{tab}
\newbox\tab@box
\let\tab@list=\empty
\newwrite\tab@write  \def\tab@file{\tab@file\tab@write0}

\outer\def\taball{\tab@init A}
\outer\def\tabchap{\tab@init C}
\outer\def\tabsect{\tab@init S}
\def\tab@init{\if@t\tab@acs\tab@pref\tab@reset\tab@@reset\tabz@}
\def\tabz@{\gz@\tabn@m}

\outer\def\tabpage{\glet\tab@page T}
\outer\def\notabpage{\glet\tab@page F}

\def\tab{\bf@t\tab@\@tab}
\def\TAB{\bf@t\tab@\@TAB}
\def\ttab{\bf@t\ttab@\@tab}
\def\TTAB{\bf@t\ttab@\@TAB}

\def\@tab{\ef@t\tabn@m\tab@pref\relax}
\def\@TAB#1{\ef@t\tabn@m\tab@pref{\def@name\t@names#1{\thef@tn@m}}}

\def\tab@{\@store1{\thef@tn@m .}}
\def\ttab@{\@sstore1\tabpref{\thef@tn@m}}
\def\tabadd{\@add1}

\def\qutab{\case@abbr\tababbr\num@lett\quote@all}
\def\qutabs{\case@abbr\tabsabbr\num@lett\quote@all}

\outer\def\tabout{\@out1\tab@page\tabpref\tabhead\emptyt@ks}
\outer\def\tabkill{\@kill1}
\outer\def\restoretab#1{\read@store{\@store1{#1}}}
\outer\def\tabrestore{\all@restore\t@names}

\outer\def\TABext{\@ext1\TAB@ext}
\def\TAB@ext{\@@ext\@TAB\@extf@t}


\newskip\htabskip   \htabskip=1em plus 2em minus .5em
\newdimen\vtabskip  \vtabskip=2.5pt
\newbox\tab@top   \newbox\tab@bot

\let\@hrule=\hrule
\let\@halign=\halign
\let\@valign=\valign
\let\@span=\span
\let\@omit=\omit

\def\@@span{\@span\@omit\@span}
\def\@@@span{\@span\@omit\@@span}

\def\sp@n{\span\@omit\advance\mscount\m@ne} 

\def\table#1#{\vbox\bgroup\offinterlineskip
  \toks@ii{#1\bgroup \unhcopy\tab@top \unhcopy\tab@bot
    ##}\afterassignment\tab@preamble \@eat}


\def\tab@preamble#1\cr{\let\tab@@vrule\tab@repeat
  \let\tab@amp@\empty \let\tab@amp\empty \let\span@\@@span
  \toks@{\tab@space#1&\cr}\the\toks@}
\def\tab@space{\tab@test{ }{}\tab@vrule}                 
\def\tab@vrule{\tab@test\vrule{\tab@add\vrule}
  \tab@@vrule}
\def\tab@repeat{\tab@test&{\tab@add&
    \let\tab@@vrule\tab@template \let\tab@amp@\tab@@amp
    \let\tab@amp&\let\span@\@@@span}\tab@template}
\def\tab@template#1&{\tab@add{\@span\tab@amp@
    \tabskip\htabskip&\tab@setup#1&\tabskip\z@skip##}
  \tab@test\cr{\let\tab@@vrule\tab@exec}\tab@space}      
\def\tab@@amp{&##}

\def\tab@test#1#2#3{\let\tab@comp= #1\toks@{#2}\let\tab@go#3%
  \futurelet\n@xt \tab@@test}
\def\tab@@test{\ifx \tab@comp\n@xt \the\toks@
    \afterassignment\tab@go \expandafter\@eat \else
  \expandafter\tab@go \fi}

\def\tab@add#1{\toks@ii\expandafter{\the\toks@ii#1}}


\def\tab@exec{\tab@r@set\everycr{\tab@body}%
  \def\halign{\tab@r@set \halign}\def\valign{\tab@r@set \valign}%
  \def\omit{\@omit \tab@setup}%
  \def\n@xt##1{\hbox{\dimen@\ht\strutbox\dimen@ii\dp\strutbox
    \advance##1\vtabskip \vrule \@height\dimen@ \@depth\dimen@ii
    \@width\z@}}%
  \setbox\tab@top\n@xt\dimen@ \setbox\tab@bot\n@xt\dimen@ii
  \def\ml##1{\relax                                      
    \ifmmode \let\@ml\empty \else \let\@ml$\fi           
    \@ml\vcenter{\hbox\bgroup\unhcopy\tab@top            
    ##1\unhcopy\tab@bot\egroup}\@ml}%
  \def\nl{\egroup\hbox\bgroup\strut}
  \tabskip\z@skip\@halign\the\toks@ii\cr}

\def\tab@r@set{\let\cr\endline \everycr\emptyt@ks
  \let\halign\@halign \let\valign\@valign
  \let\span\@span \let\omit\@omit}
\def\tab@setup{\relax \iffalse {\fi \let\span\span@ \iffalse }\fi}


\def\tab@body{\noalign\bgroup \tab@@body}           
\def\tab@@body{\futurelet\n@xt \tab@end}            
\def\tab@end{\ifcat\egroup\noexpand\n@xt
    \expandafter\egroup \expandafter\egroup \else        
  \expandafter\tab@blank \fi}
\def\tab@blank{\ifcat\space\noexpand\n@xt
    \afterassignment\tab@@body \expandafter\@eat \else   
  \expandafter\tab@hrule \fi}
\def\tab@hrule{\ifx\hrule\n@xt
    \def\hrule{\@hrule\egroup \tab@body}\else            
  \expandafter\tab@noalign \fi}
\def\tab@noalign{\ifx\noalign\n@xt
    \aftergroup\tab@body \expandafter\@eat@ \else        
  \expandafter\tab@row \fi}
\def\tab@row#1\cr{\toks@ii{\toks@ii{\egroup}
    \tab@item#1&\cr}\the\toks@ii}
\def\tab@item#1&{\tab@add{\tab@amp&#1&}
  \futurelet\n@xt \tab@cr}
\def\tab@cr{\ifx\cr\n@xt
    \expandafter\the\expandafter\toks@ii \else           
  \expandafter\tab@item \fi}

\message{references,}


\newtoks\r@names   \r@names={}
\newcount\refn@m   \refn@m=0
\newcount\ref@temp
\def\ref@type{ref}
\newbox\ref@box
\let\ref@list=\empty
\newwrite\ref@write  \def\ref@file{\ref@file\ref@write0}

\begingroup \let\refnam=\relax  
  \newhelp\ref@help{The option \string\refnam\space allows predefined
    references only and is incompatible with \string\qref(s).
    Your request will be ignored.}
  \global\ref@help=\ref@help 
  \gdef\ref@err{{\errhelp\ref@help \errmessage{Invalid request}}}
\endgroup

\begingroup
  \let\refsup=\relax \let\refsqb=\relax  
  \let\refnam=\relax                     
  \gdef\ref@setup{%
    \glet\refsup\undefined
    \if S\ref@sbn \glet\refsup\empty \glet\therefn@m\suprefn@m \fi
    \glet\refsqb\undefined
    \if B\ref@sbn \glet\refsqb\empty \glet\therefn@m\sqbrefn@m \fi
    \glet\refnam\undefined
    \if N\ref@sbn \glet\refnam\empty \glet\the@quref\nam@quref
      \glet\@@ref\ref@err \gdef\qref{\ref@err \quref}\glet\qrefs\qref
      \glet\RF@def@\RF@def@nam
      \glet\RF@find\undefined \glet\RF@search\undefined
      \glet\RF@locate\undefined \glet\@RFread\undefined
      \glet\qurefsup\ref@err \glet\sup@quref\undefined
      \glet\qurefsqb\ref@err \glet\sqb@quref\undefined
      \glet\qurefnum\ref@err \glet\num@quref\undefined
      \glet\ref@restore\ref@err
      \else \gdef\@@ref{\@store2\therefn@m}%
      \gdef\qref{\case@abbr\refabbr\num@lett\quote@all}%
      \gdef\qrefs{\case@abbr\refsabbr\num@lett\quote@all}%
      \glet\RF@def@\RF@def@num \glet\RF@print\undefined
      \gdef\ref@restore{\all@restore\r@names}\fi
    \glet\nam@quref\undefined
    \gdef\quref{\ref@unskip \num@lett\the@quref}%
    \glet\RF@def@num\undefined \glet\RF@def@nam\undefined
    \gdef\RF@restore{\all@restore\R@names}%
    \glet\ref@setup\undefined}
\endgroup

\outer\def\refsup{\glet\ref@sbn S\global\qurefsup}
\outer\def\refsqb{\glet\ref@sbn B\global\qurefsqb}
\outer\def\refnam{\if B\store@blf \opt@err \else \glet\ref@sbn N\fi}

\def\qurefsup{\let\the@quref\sup@quref}
\def\qurefsqb{\let\the@quref\sqb@quref}
\def\qurefnum{\let\the@quref\num@quref}

\outer\def\refpage{\glet\ref@page T}
\outer\def\norefpage{\glet\ref@page F}

\outer\def\refkeep{\glet\ref@kc K}
\outer\def\refclear{\glet\ref@kc C}

\def\ref{\ref@advance \refend \@ref}
\def\REF{\num@lett\@REF}
\def\@REF#1{\ref@name#1\@ref}
\def\refend{\quref{\the\refn@m}}

\def\refs{\ref@advance \ref@temp\refn@m \@ref}
\def\REFS{\num@lett\@REFS}
\def\@REFS#1{\ref@name#1\ref@temp\refn@m \@ref}
\def\refscon{\ref@advance \@ref}

\def\refsend{\quref{\the\ref@temp -\the\refn@m}}

\def\ref@advance{\ref@unskip \g@ne\refn@m}
\def\ref@name{\ref@@name\r@names}
\def\ref@@name#1#2{\ref@advance \def@name#1#2{\the\refn@m}}
\def\ref@unskip{\ifhmode \unskip \fi}

\def\suprefn@m{\the\refn@m .}
\def\sqbrefn@m{$\lbrack \the\refn@m \rbrack$}
\def\@ref{\read@store\@@ref}
\def\@@ref{\ref@setup \@@ref}
\def\refadd{\@add2}

\def\sup@quref#1{\leavevmode \nobreak \quote@all{^{#1}}}
\def\sqb@quref#1{\ \quote@all{\lbrack #1\rbrack}}
\def\num@quref{\ \quote@all}
\def\nam@quref{\@use}
\def\quref{\ref@setup \quref}
\def\qref{\ref@setup \qref}
\def\qrefs{\ref@setup \qrefs}

\outer\def\refout{{\if K\ref@kc \@out2\ref@page\refpref\refhead\emptyt@ks
  \else \@out2\ref@page\refpref\refhead\r@names
  \let\\\@RFdef \the\R@names \global\R@names\emptyt@ks
  \if L\RF@lfe \else \glet\RF@list\empty \gz@\RF@high \fi
  \gz@\refn@m \fi}}
\outer\def\refkill{\@kill2}
\outer\def\restoreref#1{\read@@store{\@store2{#1}}}
\outer\def\refrestore{\ref@restore}
\outer\def\RFrestore{\RF@restore}
\def\ref@restore{\ref@setup \ref@restore}
\def\RF@restore{\ref@setup \RF@restore}

\outer\def\REFext{\@ext2\REF@ext}
\def\REF@ext{\@@ext\ref@name\refn@m}


\newtoks\R@names   \R@names={}
\newcount\RFn@m   \newcount\RF@high
\newcount\RFmax   \RFmax=50  
\def\RF@type{RF}
\let\RF@list=\empty
\newwrite\RF@write  \def\RF@file{\RF@file\RF@write0}
\let\RF@noc=N

\begingroup \let\storefile=\relax        
  \let\RFlist=\relax \let\RFfile=\relax  
  \let\RFext=\relax \let\RF=\relax       
  \gdef\@RF{%
    \glet\RFlist\undefined
    \if L\RF@lfe \glet\RFlist\empty \glet\@RF\@RFlist
      \gdef\RF@input{\RF@list}%
      \else \glet\@RFlist\undefined \fi
    \glet\RFfile\undefined
    \if F\RF@lfe \glet\RFfile\empty \glet\@RF\@RFfile
      \else \RF@setup
      \glet\@RFfile\undefined \glet\@RFcopy\undefined
      \glet\RF@store\undefined \glet\RF@copy\undefined
      \glet\RF@@input\undefined \fi
    \glet\RFext\undefined
    \if E\RF@lfe \gdef\RFext##1 {}\glet\@RF\@RFext
      \else \glet\@RFext\undefined \fi
    \glet\RF@setup\undefined \@RF}
  \gdef\RF@setup{\ifx \storefile\undefined \glet\file@store\undefined
    \glet\file@open\undefined \glet\file@close\undefined
    \glet\file@wlog\undefined \glet\file@free\undefined
    \glet\file@copy\undefined \glet\file@read\undefined \fi}
\endgroup

\outer\def\RFlist{\glet\RF@lfe L}
\outer\def\RFfile{\if B\store@blf \opt@err \else \glet\RF@lfe F\fi}
\outer\def\RFext#1 {\if B\store@blf \opt@err \else
  \glet\RF@lfe E\gdef\RF@input{\input#1 }\RF@input \fi}

\def\@RFdef#1{\gdef#1{\RF@def#1}}
\def\@RF@list#1{\toks@\expandafter{\RF@list\RF@#1}%
  \xdef\RF@list{\the\toks@ {\the\toks@store}}}
\def\@RFcopy#1{\RF@store{\noexpand#1}}
\def\RF@store{\let\@type\RF@type \if C\RF@noc \glet\RF@noc N%
  \RF@copy \fi \glet\RF@noc O\expandafter\file@store\RF@file}
\def\RF@input{\expandafter\RF@@input\RF@file}
\begingroup \let\RF=\relax  
  \gdef\RF@copy{{\let\\\RF \let\@RF\@RFcopy
    \expandafter\file@copy\RF@file}}
  \gdef\RF@@input#1#2#3{\if O\RF@noc \let\@type\RF@type
    \file@close#2#3\glet\RF@noc C\fi \let\\\RF \file@read#3}
\endgroup

\def\RF@def#1{\ref@@name\R@names#1\RF@def@ #1}
\def\RF@def@{\ref@setup \RF@def@}
\def\RF@def@num{\toks@store\expandafter{\expandafter\RF@find
  \expandafter{\the\refn@m}}\@@ref}
\def\RF@def@nam{\ifnum\refn@m=\@ne \refadd{\RF@print}\fi}
\def\RF@test#1{\z@}

\def\RF@print{\let\@RF\@RF@print \let\RF@def\RF@test
  \let\RF@first T\RF@input}
\def\@RF@print#1{\ifnum#1>\z@
  \if\RF@first T\let\RF@first F\setbox\z@\lastbox
  \else \form@t\f@rmat \noindent \strut \fi
  \hangindent\namrefindent \the\toks@store \fi}

\def\RF@find#1{\RFn@m#1\bgroup \let\RF@def\RF@test
  \if L\RF@lfe \else \ifnum\RFn@m<\RF@high \else \RF@search \fi \fi
  \let\RF@\RF@locate \RF@list \egroup}

\def\RF@search{\global\RF@high\RFn@m \global\advance\RF@high\RFmax
  \glet\RF@list\empty \let\@RF\@RFread \let\par\relax \RF@input}
\def\@RFread#1{\ifnum#1<\RFn@m \else \ifnum#1<\RF@high
  \@RF@list#1\fi \fi}
\def\RF@locate#1#2{\ifnum#1=\RFn@m #2\fi}

\def\@RFlist#1{\@RFdef#1\@RF@list#1}
\def\@RFfile#1{\@RFdef#1\@RFcopy#1}
\let\@RFext=\@RFdef

\outer\def\RF{\num@lett\RF@}
\def\RF@#1{\ref@unskip \read@store{\@RF#1}}


\outer\def\yearpage{\glet\yearpage@yp Y}
\outer\def\pageyear{\glet\yearpage@yp P}

\def\journal#1{{\journalstyle{#1}}\j@urnal{}}
\def\journalp#1{{\journalstyle{#1}}\j@urnal}
\def\journalf#1#2#3({{\journalstyle{#1}}\j@urnal{#3}#2(}

\def\j@urnal#1#2(#3)#4*{\unskip
  \ {\volumestyle{#1\ifx @#1@\else\ifx @#2@\else
  \kern.2em\fi \fi#2}}\unskip
  \ifx @#3@\else\ifx @#4@ (#3)\else\if Y\yearpage@yp\ (#3) #4\else
  , #4 (#3)\fi \fi \fi}

\def\Phys{Phys.\ }
\def\Rev{Rev.\ }

\def\PRD{\journalp{\Phys\Rev}D}


\message{table of contents,}


\newcount\tocn@m   \tocn@m=0
\newcount\auto@toc   \auto@toc=-1
\let\toc@saved\empty

\def\toc@type{toc}
\newbox\toc@box
\let\toc@list=\empty
\newwrite\toc@write  \def\toc@file{\toc@file\toc@write0}

\outer\def\tocpage{\glet\toc@page T}
\outer\def\notocpage{\glet\toc@page F}

\outer\def\tocnone{\gm@ne\auto@toc}
\outer\def\tocchap{\gz@\auto@toc}
\outer\def\tocsect{\global\auto@toc\@ne}

\def\toc#1{\read@store{\@toc{#1}}}
\def\tocadd{\@add3}
\def\@toc{\g@ne\tocn@m
  \expandafter\@@toc\csname toc@\romannumeral\tocn@m\endcsname}
\def\@@toc#1{\pagelabel#1%
  \toks@store\expandafter{\the\toks@store\toc@fill#1}\@store3}
\def\toc@fill{\rightskip4em\@plus1em\@minus1em\parfillskip-\rightskip
  \unskip\vadjust{}\leaders\hbox to1em{\hss.\hss}\hfil}

\outer\def\tocout{\@out3\toc@page\tocpref\tochead\emptyt@ks}
\outer\def\tockill{\@kill3}
\outer\def\restoretoc#1{\read@@store{\@store3{#1}}}

\message{footnotes,}


\newcount\footn@m   \footn@m=0
\def\foot@type{foot}
\newbox\foot@box
\let\foot@list=\empty
\newwrite\foot@write  \def\foot@file{\foot@file\foot@write0}

\outer\def\footsqb{\glet\foot@bp B\glet\thefootn@m\sqbfootn@m}
\outer\def\footpar{\glet\foot@bp P\glet\thefootn@m\parfootn@m}

\outer\def\footbot{\glet\foot@be B\glet\vfootnote\vfootn@te}
\outer\def\footend{\glet\foot@be E\glet\vfootnote\foot@store}

\outer\def\footpage{\glet\foot@page T}
\outer\def\nofootpage{\glet\foot@page F}

\def\sqbfootn@m{\lbrack \the\footn@m \rbrack}
\def\parfootn@m{\the\footn@m )}

\def\foot{\hfoot \vfootnote\footid}
\def\hfoot{\g@ne\footn@m \edef\n@xt{{$^{\thefootn@m}$}}%
  \expandafter\hfootnote\n@xt}
\def\footnote#1{\hfootnote{#1}\vfootnote\footid}
\def\hfootnote#1{\let\@sf\empty
  \ifhmode\unskip\edef\@sf{\spacefactor\the\spacefactor}\/\fi
  #1\@sf \gdef\footid{#1}}
\def\vfootn@te#1{\insert\footins\bgroup \foot@style
  \llap{#1}\after@arg\@foot}
\let\fo@t=\undefined
\let\f@@t=\undefined
\let\f@t=\undefined

\def\foot@style{\footstyle  
  \interlinepenalty\interfootnotelinepenalty
  \baselineskip\footnotebaselineskip
  \splittopskip\interfootnoteskip 
  \splitmaxdepth\dp\strutbox \floatingpenalty\@MM
  \leftskip.05\hsize \rightskip\z@skip
  \spaceskip\z@skip \xspaceskip\z@skip \noindent \footstrut}

\def\foot@store#1{\read@store{\@store4{#1}}}
\def\footadd{\@add4}

\outer\def\footout{\@out4\foot@page\footpref\foothead\emptyt@ks}
\outer\def\footkill{\@kill4}
\outer\def\restorefoot#1{\read@store{\@store4{#1}}}

\def\ignorefoot{\let\foot\eat \let\hfoot\eat  
  \def\footnote{\expandafter\eat\eat}\let\hfootnote\footnote}

\message{items and points,}



\def\varitem{\afterassignment\v@ritem \setbox\z@\hbox}
\def\v@ritem{\hss \bgroup \aftergroup\v@@ritem}
\def\v@@ritem{\enskip \egroup \endgraf \noindent
  \hangindent\wd\z@ \box\z@ \ignorespaces}

\def\hvskip{\afterassignment\h@vskip \skip@}
\def\h@vskip{\unskip\nobreak \vadjust{\vskip\skip@}\lb \ignorespaces}

\def\parvskip{\bgroup \afterassignment\par@vskip \parskip}
\def\par@vskip{\parindent\hangindent \endgraf \indent \egroup
  \ignorespaces}

\def\item{\varitem to2.5em}
\def\sitem{\varitem to4.5em}
\def\ssitem{\varitem to6.5em}


\newcount\pointn@m   \pointn@m=0
\def\pointbegin{\gz@\pointn@m \point}
\def\point{\g@ne\pointn@m
  \xdef\the@label{\the\pointn@m}\item{\the@label.}}

\newcount\spointn@m   \spointn@m=96
\def\spointbegin{\global\spointn@m96 \spoint}
\def\spoint{\g@ne\spointn@m
  \xdef\the@label{\char\the\spointn@m}\sitem{(\the@label)}}

\newcount\sspointn@m   \sspointn@m=0
\def\sspointbegin{\gz@\sspointn@m \sspoint}
\def\sspoint{\g@ne\sspointn@m
  \xdef\the@label{\romannumeral\sspointn@m}\ssitem{\the@label)}}



\message{matrices and additional math symbols,}


\def\matc{\let\mat@lfil\hfil \let\mat@rfil\hfil}
\def\matl{\let\mat@lfil\relax \let\mat@rfil\hfil}
\def\matr{\let\mat@lfil\hfil \let\mat@rfil\relax}

\def\matrix#1{\null\,\vcenter{\normalbaselines\m@th
    \ialign{$\mat@lfil##\mat@rfil$&&\quad$\mat@lfil##\mat@rfil$\crcr
      \mathstrut\crcr\noalign{\kern-\baselineskip}%
      #1\crcr\mathstrut\crcr\noalign{\kern-\baselineskip}}}\,}

\def\bordermatrix#1{\begingroup \m@th
  \setbox\z@\vbox{%
    \def\cr{\crcr\noalign{\kern2\p@\glet\cr\endline}}%
    \ialign{$##\hfil$\kern2\p@\kern\p@renwd&\thinspace$\mat@lfil##%
      \mat@rfil$&&\quad$\mat@lfil##\mat@rfil$\crcr
      \omit\strut\hfil\crcr\noalign{\kern-\baselineskip}%
      #1\crcr\omit\strut\cr}}%
  \setbox\tw@\vbox{\unvcopy\z@\global\setbox\@ne\lastbox}%
  \setbox\tw@\hbox{\unhbox\@ne\unskip\global\setbox\@ne\lastbox}%
  \setbox\tw@\hbox{$\kern\wd\@ne\kern-\p@renwd\left(\kern-\wd\@ne
    \global\setbox\@ne\vbox{\box\@ne\kern2\p@}%
    \vcenter{\kern-\ht\@ne\unvbox\z@\kern-\baselineskip}\,\right)$}%
  \null\;\vbox{\kern\ht\@ne\box\tw@}\endgroup}


\mathchardef\smallsum=\dq1006
\mathchardef\smallprod=\dq1005

\def\b@mmode{\relax\ifmmode \expandafter\c@mmode \else $\fi}
\def\c@mmode#1\e@mmode{#1}
\def\e@mmode{$}
\def\defmmode#1#2{\def#1{\b@mmode#2\e@mmode}}

\defmmode\{{\lbrace}
\defmmode\}{\rbrace}

\defmmode\,{\mskip\thinmuskip}
\defmmode\>{\mskip\medmuskip}
\defmmode\;{\mskip\thickmuskip}

\defmmode{\Mit#1}{\mit#1}
\defmmode{\Cal#1}{\cal\uppercase\expandafter{#1}}

\def\dotii#1{{\mathop{#1}\limits^{\vbox to -1.4\p@{\kern-2\p@
   \hbox{\tenrm..}\vss}}}}
\def\dotiii#1{{\mathop{#1}\limits^{\vbox to -1.4\p@{\kern-2\p@
   \hbox{\tenrm...}\vss}}}}
\def\dotiv#1{{\mathop{#1}\limits^{\vbox to -1.4\p@{\kern-2\p@
   \hbox{\tenrm....}\vss}}}}

\let\barsymbol -
\mathchardef\tildesymbol=\dq0218
\def\hatsymbol{{\mathchoice{\null}{\null}{\,\,\hbox{\lower 10\p@\hbox
    {$\widehat{\null}$}}}{\,\hbox{\lower 20\p@\hbox
       {$\hat{\null}$}}}}}


\def\begin@stmt{\par\noindent\bpargroup\stmttitlestyle}
\def\adv@stmt#1#2#3#4{\begin@stmt
  \count@\ifx#2#3#1 \else\z@ \glet#2#3\fi \advance\count@\@ne
  \xdef#1{\the\count@}\edef\the@label{#4#1}}

\def\make@stmt{\ \the@label \make@@stmt}
{\catcode`\:\active
  \gdef\make@@stmt{\ \catcode`\:\active \let:\end@stmt}
}
\def\end@stmt{\catcode`\:\@ther \unskip :\stmtstyle
  \enskip \ignorespaces}

\def\defstmt#1#2#3{\expandafter\def@stmt \csname#1\endcsname
  {#2}{#1@stmt@}#3@}
\def\def@stmt#1#2#3#4#5@{\bgroup
  \toks@{\begin@stmt #2\make@@stmt}\toks@ii{#2\make@stmt}%
  \if#4n\xdef#1{\the\toks@}\else
    \xdef#1{\csname#3adv\endcsname \the\toks@ii}%
    \if#4=\edef\n@xt{\expandafter\noexpand
      \csname#5@stmt@adv\endcsname}\else
      \toks@{\empty}\toks@ii\toks@ \if#4c\toks@{\dot@pref}\fi
      \if#4s\toks@{\sect@pref}\if#5c\toks@ii{\sect@dot@pref}\fi \fi
      \if#5a\toks@ii\toks@ \fi
      \edef\n@xt{\noexpand\adv@stmt \csname#3num\endcsname
        \csname#3save\endcsname \the\toks@ \the\toks@ii}\fi
    \expandafter\glet\csname#3adv\endcsname\n@xt \fi
  \egroup}

\def\Prf{\par\noindent\bpargroup\prftitlestyle \case@language\prfhead
  \let\stmtstyle\prfstyle \make@@stmt}

\message{save, restore and start macros,}


\def\save@type{.texsave }  
\newread\test@read
\newwrite\save@write

\def\s@ve{\immediate\write\save@write}
\begingroup \catcode`\:=\active \catcode`\;=\active
  \outer\gdef\save#1 {{\let\,\space
    \immediate\openout\save@write#1\save@type
    \s@ve{;* definitions for :restore #1 \date\space- \thetime\space*}%
    \s@ve{:comment}%
    \s@ve{:mainlanguage:\case@language{german\else english}}%
    \save@page \save@chap
    \save@equ \save@fig \save@tab \save@ref \save@toc \save@foot
    \bgroup \def\n@xt##1.##2{\advance##2\@ne \s@ve{:start##1\the##2}}%
      \n@xt chap.\chapn@m \n@xt sect.\sectn@m
      \n@xt appendix.\appn@m \n@xt equ.\eqn@m
      \n@xt fig.\fign@m \n@xt tab.\tabn@m
      \n@xt ref.\refn@m \n@xt toc.\tocn@m
      \n@xt foot.\footn@m \egroup
    \s@ve{:\ifx\chap@@eq\sect@@eq app\else
      \ifnum\chapn@m=\z@ sect\else chap\fi \fi init}%
    \@save0\@save1\@save2\@save3\@save4%
    \if K\ref@kc \s@ve{:endcomment}\fi
    \save@restore ref \r@names  \save@restore RF \R@names
    \if C\ref@kc \s@ve{:endcomment}\fi
    \save@restore lbl \l@names  \save@restore eq \e@names
    \save@restore fig \f@names  \save@restore tab \t@names
    \s@ve{;*  end of definitions  *}\immediate\closeout\save@write
    }\wlog{* file #1\save@type saved *}}
  \gdef\save@restore#1 {\def\\{\s@ve{:#1restore}%
      \let\\\save@@restore \\}\the}
  \gdef\save@@restore#1{\toks@\expandafter{#1}%
    \s@ve{:dorestore\string#1{\the\toks@}}}
  \gdef\save@page{\s@ve{:\@opt\page@tbn Ttop Bbot Nno *pagenum%
      :\@opt\head@lrac Llef Rrigh Aal Ccen *thead%
      :\@opt\foot@lrac Llef Rrigh Aal Ccen *tfoot%
      :page\@opt\page@ac Aall Cchap *%
      :\@opt\ori@pl Pportrait Llandscape *}%
    \s@ve{:startpage\@opt\page@ac C\the\pageno@pref. *\the\pageno}}
  \gdef\save@chap{\s@ve{:\@opt\chap@page Fno *chappage%
      :\@opt\chap@yn Nno *chapters%
      :\@opt\chap@ar Aarabic Rroman *chapnum}}
  \gdef\save@equ{\s@ve{:equ\@opt\eq@acs Aall Cchap Ssect *%
      :equ\@opt\eq@lrn Lleft Rright Nnone *%
      :equ\@opt\eq@fs Ffull Sshort *}}
  \gdef\save@fig{\s@ve{:\@opt\fig@page Fno *figpage%
      :fig\@opt\fig@acs Aall Cchap Ssect *}}
  \gdef\save@tab{\s@ve{:\@opt\tab@page Fno *tabpage%
      :tab\@opt\tab@acs Aall Cchap Ssect *}}
  \gdef\save@ref{\s@ve{:\@opt\ref@page Fno *refpage%
      :ref\@opt\ref@kc Kkeep Cclear *%
      :ref\@opt\ref@sbn Ssup Bsqb Nnam *%
      :\@opt\yearpage@yp Yyearpage Ppageyear *}}
  \gdef\save@toc{\s@ve{:\@opt\toc@page Fno *tocpage%
      :toc\ifcase\save@@toc \else none\fi}}
  \gdef\save@foot{\s@ve{:\@opt\foot@page Fno *footpage%
      :foot\@opt\foot@be Bbot Eend *%
      :foot\@opt\foot@bp Bsqb Ppar *}}
\endgroup
\def\@opt#1#2#3 #4{\if #1#2#3\fi \if #4*\else
  \expandafter\@opt\expandafter#1\expandafter#4\fi}
\def\save@@toc{\auto@toc chap\or sect}

\newif\ifcr@ss
\begingroup \let\comment=\relax
  \outer\gdef\restore{\@kill0\@kill1\@kill2\@kill3\@kill4%
    {\def\\##1{\glet##1\undefined}%
      \def\n@xt##1{\the##1\global##1\emptyt@ks}%
      \n@xt\l@names \n@xt\e@names \n@xt\f@names \n@xt\t@names
      \n@xt\r@names \let\\\@RFdef \n@xt\R@names}
    \bgroup \let\comment\relax \let\d@rest@re\dorest@re \@restore}
\endgroup
\def\crossrestore#1 {\bgroup \openin\test@read#1\save@type
  \ifeof\test@read \let\n@xt\egroup
    \message{* file #1\save@type missing *}%
  \else \closein\test@read
    \def\n@xt{\@restore#1 }\let\d@rest@re\docr@ss \fi \n@xt}
\def\@restore#1 {\input #1\save@type \egroup}
\def\all@restore{\glet\name@list}
\outer\def\dorestore{\bgroup \catcode`\@\l@tter \num@lett\d@restore}
\def\d@restore#1#2{\egroup \toks@{#2}\d@rest@re#1}
\def\dorest@re#1{\def@name\name@list#1{\the\toks@}}
\def\docr@ss#1{\ifx#1\undefined \cr@sstrue
    \else \expandafter\testcr@ss#1\cr@ss\@@ \fi
  \ifcr@ss \expandafter\testcr@ss\the\toks@\cr@ss\@@ \else \cr@sstrue \fi
  \ifcr@ss\else \expandafter\dorest@re\expandafter#1\fi}
\def\testcr@ss#1\cr@ss#2\@@{\ifx @#2@\toks@{\cr@ss#1}\cr@ssfalse
  \else \cr@sstrue \fi}
\let\cr@ss=\empty

\def\def@name#1#2{\let\name@list#1\add@name#2\xdef#2}
\def\del@name#1{\bgroup
  \def\n@xt##1\\#1##2\\#1##3\@@##4{\global##4{##1##2}}%
  \def\del@@name##1{\expandafter\n@xt\the##1\\#1\\#1\@@##1}%
  \del@@name\l@names \del@@name\e@names \del@@name\f@names
  \del@@name\t@names \del@@name\r@names \del@@name\R@names \egroup}
\def\add@name#1{\del@name#1%
  {\global\name@list\expandafter{\the\name@list\\#1}}}

\def\kill{\num@lett\@k@ll}
\def\@k@ll#1{\def\k@ll##1{\ifx##1\k@ll \let\k@ll\relax \else
    \del@name##1\glet##1\undefined \fi \k@ll}\k@ll#1\k@ll}


\outer\def\startchap{\st@rt\chapn@m}
\outer\def\startsect{\st@rt\sectn@m}
\outer\def\startappendix{\st@rt\appn@m}
\outer\def\startequ{\st@rt\eqn@m}
\outer\def\startfig{\st@rt\fign@m}
\outer\def\starttab{\st@rt\tabn@m}
\outer\def\startref{\st@rt\refn@m}
\outer\def\starttoc{\st@rt\tocn@m}
\outer\def\startfoot{\st@rt\footn@m}

\def\st@rt#1{\gm@ne#1 \global\advance#1}

\message{installation dependent parameters,}




\hoffset@corr@p=86mm   \voffset@corr@p=98mm      
\hoffset@corrm@p=-5mm   \voffset@corrm@p=4mm      
\hoffset@corr@l=134mm   \voffset@corr@l=70mm     
\hoffset@corrm@l=-5mm   \voffset@corrm@l=-4mm     

\catcode`\@=12 

\message{and default options.}



\hbadness=2000  

\newsect        

\mainlanguage\english 

\botpagenum     
\centhead       
\centfoot       
\pageall        
\portrait       
\titlepage      
\nochappage     
\arabicchapnum  
\chapters       
\equchap        
\equshort       
\equright       
\storelist      
\figall         
\figpage        
\nographics     
\taball         
\tabpage        
\refsup         
\refpage        
\refkeep        
\RFlist         
\yearpage       
\tocpage        
\tocnone        
\footsqb        
\footbot        
\footpage       
\matc           

\wlog{summary of allocations:}
\wlog{last count=\number\count10 }
\wlog{last dimen=\number\count11 }
\wlog{last skip=\number\count12 }
\wlog{last muskip=\number\count13 }
\wlog{last box=\number\count14 }
\wlog{last toks=\number\count15 }
\wlog{last read=\number\count16 }
\wlog{last write=\number\count17 }
\wlog{last fam=\number\count18 }
\wlog{last language=\number\count19 }
\wlog{last insert=\number\count20 }

   \def\Fmtversion{2.0}

\edef\fmtversion{\Fmtversion(\fmtname\space\fmtversion)}
\let\fmtname=\Fmtname 
\everyjob={
  \immediate\write16{\fmtname\space version \fmtversion\space
    format preloaded.}%
  \input phystime      
  \input physupdt }    
\immediate\write16{Version \fmtversion\space format loaded.}%


\def\wlog#1{} 
\catcode`\@=11


\def\({\relax\ifmmode[\else$[$\nobreak\hskip.3em\fi}
\def\){\relax\ifmmode]\else\nobreak\hskip.2em$]$\fi}

\def\gappr{\mathpalette\under@rel{>\approx}}
\def\lappr{\mathpalette\under@rel{<\approx}}
\def\gsim{\mathpalette\under@rel{>\sim}}
\def\lsim{\mathpalette\under@rel{<\sim}}
\def\under@rel#1#2{\under@@rel#1#2}

\def\under@@rel#1#2#3{\mathrel{\mathop{#1#2}\limits_{#1#3}}}

\def\under@@rel#1#2#3{\mathrel{\vcenter{\hbox{$%
  \lower3.8pt\hbox{$#1#2$}\atop{\raise1.8pt\hbox{$#1#3$}}%
  $}}}}

\def\widebar#1{\mkern1.5mu\overline{\mkern-1.5mu#1\mkern-1.mu}\mkern1.mu}
\def\parenbar{\mathpalette\p@renb@r}
\def\p@renb@r#1#2{\vbox{%
  \ifx#1\scriptscriptstyle \dimen@.7em\dimen@ii.2em\else
  \ifx#1\scriptstyle \dimen@.8em\dimen@ii.25em\else
  \dimen@1em\dimen@ii.4em\fi\fi \offinterlineskip
  \ialign{\hfill##\hfill\cr
    \vbox{\hrule width\dimen@ii}\cr
    \noalign{\vskip-.3ex}%
    \hbox to\dimen@{$\mathchar300\hfil\mathchar301$}\cr
    \noalign{\vskip-.3ex}%
    $#1#2$\cr}}}


\def\mppae@text{{Max-Planck-Institut f\"ur Physik}}
\def\mppwh@text{{Werner-Heisenberg-Institut}}

\def\mppaddresstext{Postfach 40 12 12, D-8000 M\"unchen 40\else
  P.O.Box 40 12 12, Munich (Fed.^^>Rep.^^>Germany)}

\def\mppaddress{\address{\mppae@text \nl -- \mppwh@text\space --\nl
  \case@language\mppaddresstext}}

\def\mppnum#1{\topright{MPI-Ph/#1}}


\font\fourteenssb=cmssdc10 scaled \magstep2 
\font\seventeenssb=cmssdc10 scaled \magstep3 

\def\letter#1#2{\b@lett@r{26}%
  \centerline{\seventeenssb \uppercase\mppae@text}%
  \centerline{\fourteenssb \uppercase\mppwh@text}%
  \centerline{\strut#1}\vskip.5cm%
  \e@lett@r{\hss\vtop to5cm{\hsize55mm%
    \lftline{\strut}\eightrm  \setbaselineskip=12pt \vfil
    \lftline{F\"OHRINGER RING 6}\lftline{\tenrm D-8000 M\"UNCHEN 40}%
    \lftline{\case@language{TELEFON\else PHONE}: (089) 3 23 08
      \if!#2!\else - #2 \case@language{oder\else or} \fi-1}%
    \lftline{TELEGRAMM:}\lftline{PHYSIKPLANCK M\"UNCHEN}%
    \lftline{TELEX: 5 21 56 19 mppa d}%
    \lftline{TELEFAX: (089) 3 22 67 04}%
    \lftline{POSTFACH 40 12 12}
    \ifx\EARN\undefined\else\vskip5\p@\lftline{EARN/BITNET: \EARN
      @DM0MPI11}\fi \vfil}}}

\def\b@lett@r#1{\endpage \begingroup \doublespace \vglue-#1mm}

\def\e@lett@r#1#2{\skippagenum T\skipheadline T\skipfootline T%
  \line{\vtop to47mm{\lftline{\llap{\vbox to\z@{\vskip171\p@
      \hrule\@width7\p@\vss}\hskip57\p@}\strut}\vskip2mm\vfil
    \addressspacing \dimen@\baselineskip \dimen@ii-2.79ex%
    \advance\dimen@ii\dimen@ \baselineskip\dimen@\@minus\dimen@ii
    \let\nl\cr\use@nl \halign{##\hfil\crcr#2\crcr}\vfil}#1}%
  \vskip1cm\rtline{\thedate}\vskip1cm\@plus1cm\@minus.5cm\endgroup}

\let\addressspacing=\empty


\def\myname{Dr.\ Xxxx Xxxxxxxxxx\nl Physiker}
\def\myaddress{Xxxxxxx Stra\ss e  ??\nl
    \llap{D--8000\quad}M\"unchen ??\nl
    Tel:\ (089) \vtop{\hbox{?? ?? ?? (privat)}%
                      \hbox{3 18 93-??? (B\"uro)}}}
\def\myletter{\b@lett@r{26}\line{\let\nl\cr \use@nl \caps
  \vtop to25mm{\halign{\strut##\hfil\crcr\myname\crcr}\vfil}\hfil
  \vtop to25mm{\tenpoint
    \halign{\strut##\hfil\crcr\myaddress\crcr}\vfil}}%
  \e@lett@r\empty}


\def\firstpageoutput{\physoutput
  \global\output{\setbox\z@\box@cclv \deadcycles\z@}}


\def\veq{\afterassignment\v@eq \dimen@}
\def\v@eq{$$\vcenter to\dimen@{}$$}

\def\veqn{\afterassignment\v@eqn \dimen@}
\def\v@eqn{$$\vcenter to\dimen@{}\eqn$$}

\def\heq{\afterassignment\h@eq \dimen@}
\def\h@eq{$\hbox to\dimen@{}$ }

\def\wlog{\immediate\write\m@ne} 
\catcode`\@=12 

\def\wlog#1{} 
\catcode`\@=11

\outer\def\pthnum#1{\errmessage{***** \string\pthnum\space is no longer
         supported, use \string\mppnum\space instead}}

\def\wlog{\immediate\write\m@ne} 
\catcode`\@=12 



  \overfullrule=0pt
  \parindent=0pt
  \equfull
  \footpar \refsqb
  \crossrestore{strga}
\def\l{\lambda}
\def\a{\alpha}
\def\b{\beta}
\def\d{\delta}
\def\k{\kappa}

\def\p{\partial}

\def\r{\rho}
\def\bpsi{\bar\psi}
\def\dslash{\p\llap{/}}
\def\pslash{p\llap{/}}

\def\Ga{\Gamma}
\def\Gacl{\Gamma_{cl}}
\def\ha{{1\over 2}}

\def\bW{{\bf W}}
\def\N{{\cal N}}
\def\C{{\cal C}}
\def\R{{\cal R}}
\def\ga{\gamma}

\def \footstyle{\tenpoint}
\def \G{\Gamma}
\def \Q#1#2 {Q(#1,#2)}
\def \t {\tau }
\def \frac#1#2 {\hbox{${#1\over #2}$}}
\def \bl {\b _ \l}
\def \kdk {\k \p _\k}
\def\mdm {m \p _m}
\def \te {\tau_{\scriptscriptstyle 1}}
\def \mhi {m_{\hbox{\eightpoint $H$}}}
\def \mf {m_{\hbox{\eightpoint $f$}}}

\RF\JB1{{\caps F. Jegerlehner},
       {Renormalizing the standard model,
        in: Proceedings of the 1990 Theoretical Advanced Study Institute in
        Elementary Particle Physics, Boulder, Colorado, ed. M. Cvetic and P.
        Langacker, 1991, Singapore.}}
\RF\CaSy{{\caps K. Symanzik},
         {\sl Comm. Math. Phys.} {\bf 18} (1970) 227; \lb
         {\caps K. Symanzik}, {\sl Comm. Math. Phys.} {\bf 23} (1971) 49;\lb
    {\caps C.G. Callan}, 
        {\sl Phys. Rev.} {\bf D2} (1970) 1541.}
\RF\R12{{\caps W. Zimmermann},
       {\sl Comm. Math. Phys.} {\bf 15} (1969) 208.}
\RF\R13{
        {\caps J.H. Lowenstein, W. Zimmermann},
       {\sl Comm. Math. Phys.} {\bf 44} \hbox{(1975) 73};
       {\caps J.H. Lowenstein},
       {\sl Comm. Math. Phys.} {\bf 47} (1976) 53.}
\RF\ZiRG{{\caps W. Zimmermann},
          {\sl  Comm. Math. Phys.} {\bf 76} (1980) 39.}
\RF\Bogo{{\caps N.N. Bogoliubov, D.V. Shirkov},
      Introduction to the theory of quantized fields; J.Wiley (1980).}
\RF\LowRG{{\caps J.H. Lowenstein},
                {\sl Comm. Math. Phys.} {\bf 24} (1971) 1.}
\RF\ColRen{{\caps J. Collins}, Renormalization, Cambridge University
             Press 1984.}
\RF\Gross{{\caps D. Gross}, Applications of the renormalization group,
         in: Methodes in field theory, Les Houches 1975; ed. R.Balian and
           J.Zinn-Justin, North-Holland 1976.}
\RF\BreiMai{{\caps P.~Breitenlohner, D.~Maison},
              {\sl Comm.~Math.~Phys.} {\bf 52} (1977) 11.}
\RF\CouHil{{\caps R.~Courant, D.~Hilbert}, Methoden  der Mathematischen
            Physik II, Springer Verl.~Berlin 1968; \lb
             {\caps E.~Kamke}, Differentialgleichungen
             II, Akad.~Verlagsgesellsch.~Leipzig 1959. }
\RF\CalSym{{\caps E.~Kraus},
           Callan-Symanzik and renormalization group equation in a model
           with spontaneous breaking of the symmetry;\lb
           Bern Preprint BUTP 93-10, to be published in  {\sl Z.~Phys.~C}.}
\RF\KrAs{{\caps E.~Kraus},
          Asymptotic normalization conditions and mass independent
          renormalization group functions; Bern Preprint, in preparation}
\RF\GMLRG{{\caps E.C.G.~Stueckelberg, A.~Peterman},
         {\sl Helv.~Phys.~Acta} {\bf 26} \hbox{(1953) 433};
          {\caps M.~Gell-Mann, F.E.~Low},
         {\sl Phys.~Rev.} {\bf 95} (1954) 1300.}
\RF\MarIm{{\caps W.J.~Marciano},
             \PRD20(1979)274* .}
\RF\IZU{{\caps C.~Itzykson, J.-B.~Zuber},
         Quantum field theory, McGraw-Hill 1985}
\RF\BePu{{\caps D.~Bessis, M.~Pusterla},
          \journal{Nuov. Cim} 54 (1968) 234*.}
\topright{BUTP-93/26}
\pubdate{November 93}
\title{The structure of the invariant charge  \nl in massive
             theories with one coupling}
\author{Elisabeth Kraus
        \rm\footnote
        *{Supported by  Schweizerischer Nationalfonds and
          Deutsche Forschungsgemeinschaft}}
\address{Institut f\"ur Theoretische Physik, Universit\"at Bern\nl
          Sidlerstrasse 5, CH-3012 Bern, Switzerland}
\abstract{
 Invariance under finite renormalization group (RG) transformations
is used to structure
  the invariant charge in models with one coupling in the 4 lowest orders of
perturbation theory. In every order there starts a
RG-invariant,
 which is uniquely continued to higher
orders. Whereas in massless models the RG-invariants
 are power series in logarithms,
there is no such requirement in a massive model. Only, when one applies the
  Callan-Symanzik (CS) equation of the respective theories,  the
high-energy behavior of the RG-invariants
is restricted. In models, where
the CS-equation has the same form as the RG-equation,
 the massless limit is reached smoothly, i.e.~the $\b$-functions are
constants in the asymptotic limit and the RG-functions
starting the new invariant tend to logarithms. On the other
hand in the spontaneously broken models with fermions the CS-equation
 contains a $\b$-function of  a physical mass. As a consequence
the $\b$-functions depend on the normalization point also in the asymptotic
region and a
 mass independent  limit does not exist anymore.}
\endpage

\pageno = 1
\baselineskip = 0.99 \baselineskip
\chap{Introduction} By now in all relevant theories of particle physics the
calculations of the 1-loop order are nearly completed and show -- especially
in the standard model -- a very good agreement with the experiment in this
approximation. In order to get further restrictions and information about the
reliability of the standard model one has to take into account higher order
corrections.  Whereas one starts to calculate the 2-loop order systematically,
the question arises if one could draw some general conclusions from the lowest
order to contributions appearing necessarily in higher orders. Of special
interest thereby is the dependence on the scales of the theory, which are the
physical masses and the normalization point needed in order to fix the
couplings at their experimental value.  As a natural tool suggests itself
renormalization group invariance. Except for the trivial case, where the
interactions do not depend on the momenta and are constant, as it happens in
the classical approximation, renormalization group invariance is only realized
to all orders of perturbation theory. Consequently the lowest order induces
the next one necessarily up to the addition of new renormalization group
invariants.

In quantum field theory renormalization group invariance is mostly applied in
its infinitesimal version, where it is given as a partial differential
equation, the renormalization group equation.  In this form it has
specifically been applied to deduce the high-energy behavior of the
interactions. In a 1-coupling theory the renormalization group equation
approximated with the 1-loop $\beta $-function can be solved analytically and
therefore its solution, the 1-loop invariant, is known to all orders of
perturbation theory together with its analytical continuation to a
non-perturbative regime.  The applications of the 1-loop invariant are divided
into two different classes: Originally it was used in QED \quref{\GMLRG},
where the coupling has a 1-loop $\beta $-function with a positive sign. One
can roughly estimate by considering the perturbative power series, that in
such theories the 1-loop invariant dominates all higher order contributions in
a certain range of asymptotic momenta, where perturbation theory, i.e.~the
power series expansion of the 1-loop invariant, is meaningful.  If one has
fixed the electromagnetic coupling at low momenta, with the help of the 1-loop
renormalization group invariant it can be calculated at a much higher scale,
as it is for example the mass of the Z-boson \quref{\MarIm}.  In this form it
is successfully applied also in the standard model for the coupling of the
electromagnetic interaction \quref{\JB1}.  It is this aspect of the
renormalization group which is meant by the so-called concept of improvement.
On the other hand in theories with a negative 1-loop $\b $-function, as it is
e.g.~in QCD, the analytical form of the 1-loop invariant is used to prove
asymptotic freedom in the ultraviolet, i.e.~the coupling approaches zero for
large momenta \quref{\Gross}.  It is again the 1-loop invariant which
determines the behavior of the interaction in infinity, as it can be estimated
from the renormalization group equation.  For massless theories with a
positive $\b $-function similar estimates are valid in the infrared region.
Deviations from the leading behavior are calculable in the next-to-leading
logarithms summation (cf.~\quref{\ColRen}).  These approximations are not
renormalization group invariants by themselves, because the renormalization
group equation with the 2-loop $\beta $-function added, is not analytically
solvable anymore.

In this paper we investigate a different approach to renormalization group
invariance, namely finite renormalization group transformations \quref{\Bogo ,
\ZiRG }. As a first application we analyze in 1-coupling theories the
invariant charge, which is constructed as an invariant under renormalization
group transformations.  We formulate the requirement of renormalization group
invariance order by order in perturbation theory. In contrast to the solutions
of the partial differential equation one does not get the all order summation
in one stroke, but one is able to structure the Green functions according to
their invariance under renormalization group transformations.  We show that in
every order a new renormalization group invariant starts, whose form is given
by the solution of a functional equation.  Strings of lower order induced
functions run through all orders of perturbation theory and start to be
interwoven with each other from three loop order onwards. In this form
renormalization group invariance can be applied, if one wants to know, which
terms arise necessarily in the second order once one has calculated the finite
Green functions in 1-loop order. We perform the explicit calculations for the
four lowest orders of perturbation theory.

 Not only the renormalization group invariance gives insight into the momentum
dependence of the Green functions, but also the dilatations do so.  For the
theories we consider in this paper they are too expressible as a partial
differential equation, the Callan-Symanzik equation, which is -- concerning
its form -- similar to the one of the renormalization group equation
\quref{\CaSy}.  Therefore the invariant charge which we have constructed with
the help of finite renormalization group invariance has to be a solution of
the Callan-Symanzik equation as well.

In a first stage we consider theories, where the Callan-Symanzik equation has
exactly the same form as the renormalization group equation, i.e.~they contain
both the same differential operators.  Examples for such models are the
massive $\phi ^4 $ theory, the $O(N) $-models and the purely scalar
$U(1)$-axial model with spontaneous breaking of the symmetry.  Applying the
Callan-Symanzik operator to the general renormalization group invariant
solution one can assign to any renormalization group function a well defined
high-energy behavior, especially for all these theories the massless limit is
reached smoothly in the asymptotic region. Apart from these restrictions on
the asymptotic behavior the renormalization group solution is shown to be in
complete agreement with the Callan-Symanzik equation.

In the last part we carry out the same analysis for the $U(1)$-axial model
with one fermion, which gets its mass via the spontaneous symmetry breaking.
In contrast to the models above, the Callan-Symanzik equation differs from the
renormalization group equation because a $\b $-function of a physical mass
appears \quref{\CalSym}.  As a consequence of this difference the strings of
renormalization group invariants are not separated into 1-loop, 2-loop and so
on induced contributions, but 1-loop induced contributions appear in all
renormalization group invariant functions. Moreover in the asymptotic limit,
which we consider for simplification, the massless theory is not reached
anymore. For instance the $\b $-functions of the Callan-Symanzik equation and
the renormalization group equation depend on logarithms of the normalization
point.

In section 2 of this paper we introduce the 1-coupling models and give the
1-loop invariant solution of the renormalization group equation with its full
mass dependence in the massive $\phi ^4 $-model. To finite renormalization
group transformations we turn in section 3. There we solve the four lowest
orders, determine the $\beta$-functions in terms of the renormalization group
functions and consider, how the structure is realized diagrammatically in the
2-loop order of the $\phi ^4 $-theory. In section 4 and 6 we apply the
Callan-Symanzik equation on the invariant charge, first in the pure scalar
models encountered above and then in the spontaneously broken $U(1) $-axial
model with fermions.  Section 5 contains a few comments on reparametrizations
of the coupling.  In the last section we give a short summary of the results
and an outlook to further applications and consequences.  \baselineskip = 19
pt

\chap{Renormalization group solution in the massive 1-coupling theory} For a
first application of finite renormalization group transformations we consider
simple scalar models with one coupling $\l $ and one mass parameter.  Examples
for such models are the $O(N)$-models with the classical action $$ \Gacl =
\int \ha \p {\vec \varphi} \cdot \p {\vec \varphi} - \ha m^2 {\vec
\varphi}\cdot {\vec \varphi} - {\l \over 4!} \bigl({\vec \varphi} \cdot {\vec
\varphi}\bigr)^2, \EQN{\A1} $$ where ${\vec \varphi} = ( \phi_1,\phi _2 , ...,
\phi _N) $.  But the analysis comprises also the purely scalar $U(1)$-axial
model in its spontaneously broken phase (linear $\sigma$-model with a massless
Goldstone boson): $$ \Gacl = \int \Bigl( \frac 12 \bigl(\p \phi_1 \p \phi_1 +
\p \phi_2 \p \phi_2 \bigr) - \frac 12 m^2 \phi_1^2 - \frac 12 m \sqrt{ \frac
\l 3 } \phi_1 ( \phi_1^2 + \phi_2^ 2) - \frac \l{4!} ( \phi_1^2 + \phi_2^ 2)^2
\Bigr) \EQN{\F6sc} $$ These models are distinguished also by the fact, that
one is able to derive a Callan-Symanzik (CS) equation rigorously to all orders
of perturbation theory, which contains exactly the same differential operators
as the renormalization group (RG) equation. A counterexample to these models
we analyze in section 6.

In perturbation theory the Green functions are defined according to the
Gell-Mann Low formula, by a suitable subtraction scheme and the respective
Ward identities.  The free parameters have to be fixed by appropriate
normalization conditions, which we choose for the models \queq{\A1} and
\queq{\F6sc} in the following way: $$\eqalign{ & \p_{p^2} \Gamma_2
(p^2)\bigg|_ {p^2 = \kappa ^2} = 1 \qquad \qquad \qquad \Gamma _2 (p^2)\bigg|
_{p^2 = m^2} = 0 \cr & \G _4 (p_1, p_2, p_3, p_4) \bigg|_{{ p_i^2 = \kappa^2
\hfill} \atop { p_i p_j = - {\kappa^2 \over 3}} } = - \lambda , } \EQN{\A2} $$
where we have defined $ \G _2 = \G_ {\phi _1 \phi _1 }, \G_4 = \G_ {\phi_1
\phi _1 \phi _1 \phi _1 } $. For the spontaneously broken model \queq{\A2} has
to be enlarged by the requirement, that the vacuum expectation value is
vanishing: $$ \G _{\phi_1} = 0 \SUBEQN{\A2a} $$ Throughout the paper we
restrict ourselves to on-shell normalization for the mass in order to be able
to exploit order by order finite renormalization group invariance.  The
normalization point for the coupling and the wave function is taken to be in
the Euclidean region ($\kappa^2 < 0$).

The behavior of the Green functions under an infinitesimal change of the
normalization point $\k$ is expressed by the RG-equation $$ \bigl( \kappa \p_
\kappa +\tilde \b _\l \p _\l - \tilde \ga \N \bigr) \G (\phi_i) = 0 \quad
\hbox{with} \quad \N= \sum_{i=1}^N \int \phi_i {\d \over \d \phi_i}.
\EQN{\A4} $$ The CS-equation describes the breaking of the dilatations by the
mass term and the dilatational anomalies represented by the function $\b_\l$
and the anomalous dimension $\ga$. For the models above it has the general
form: $$\bigl( m \p _m + \kappa \p _ \kappa +\b _\l \p _\l - \ga \N \bigr) \G
(\phi_i)= \( \Delta_m \)^2 _2 \cdot \G (\phi_i) \EQN{\A3} $$ The right-hand
side is constructed to behave as a truly soft insertion, i.e.~it vanishes for
large non-exceptional momenta.  In the O(N)-models it is just given by the
soft mass insertion $$ \Delta_m = \a \int ( -m^2 \vec \varphi\cdot \vec
\varphi ) \SUBEQNBEGIN{\A3a} $$ whereas in the spontaneously broken models the
construction is much more subtle (cf.~section~6 and \quref{\CalSym}): $$
\Delta _m = \int ( - m^2 \phi_1 ^2 - \frac 12 m \sqrt{ \frac \l 3 } \phi_1
(\phi _1 ^2 + \phi _2^2) + O(\hbar) ) \SUBEQN{\A3b} $$

In this paper we restrict the considerations concerning the RG-transformations
to the invariant charge defined as a combination of perturbatively constructed
Green functions $$\eqalign{ & Q(p^2, m^2, \k ^2, \l) = \cr & \, \, \, - \Ga_4
(p_1,p_2,p_3,p_4,m^2,\k^2,\l) \prod_{k=1}^4 \biggl( \p _{p_k^2} \Ga_2
(p_k^2,m^2,\k^2, \l) \biggr)^{-\ha}\Bigg|_{{p_i^2 = p^2} \hfill \atop { p_i
p_j = - {p^2 \over 3} }} } \EQN{\DEFQ} $$ where again $p^2 < 0 $ for
definiteness.  According to \queq{\A2} it has well-defined normalization
properties $$ Q(p^2 , m^2 , \k^2 ,\l ) \Big| _{p^2 = \k^2 } = \l .
\EQN{\NORMQ} $$ Furthermore it is dimensionless $$ (p^2 \p _{p^2} + m^2 \p
_{m^2} + \k^2 \p _{\k^2}) Q(p^2, m^2, \k ^2, \l) = 0 \EQN{\A6} $$ and depends
therefore only on the dimensionless ratios $ { p^2 \over \k^2} $ and $ { m^2
\over p^2} $.  $$ Q(p^2, m^2, \k ^2, \l) = Q(\hbox{${p^2\over
\k^2}$},\hbox{${m^2 \over p^2}$} , \l) \EQN{\A7} $$ The most important
property of the invariant charge concerning the RG is the fact, that it is an
invariant under the RG-transformations.  For this reason it satisfies the
homogeneous RG-equation $$ \bigl( \kappa \p _ \kappa +\tilde \b _\l \p _\l
\bigr) Q(\hbox{${p^2\over \k^2}$},\hbox{${m^2 \over p^2}$} , \l) = 0
\EQN{\RGQ} $$ and the CS-equation without anomalous dimensions: $$ \bigl( m \p
_m + \kappa \p _\kappa +\b _\l \p _\l ) Q(\hbox{${p^2\over \k^2}$},\hbox{${m^2
\over p^2}$} , \l) = Q_m (\hbox{${p^2\over \k^2}$},\hbox{${m^2 \over p^2}$} ,
\l) \EQN{\CSQ} $$ with $$ Q_m (\hbox{${p^2\over \k^2}$},\hbox{${m^2 \over
p^2}$} , \l) = \Bigl( \bigl( \( \Delta_m \)^2_2 \cdot \G \bigr) _4 \bigl( \p
_{p^2} \G _2 \bigr) ^{-2} - 2 \p _{p^2}\bigl( \( \Delta _m \)^2_2 \cdot \G
\bigr)_2 \G _4 \Bigr) \SUBEQNBEGIN{\CSQa} $$ all functions understood to be
perturbatively expanded.

Before we turn to the finite RG-transformations we want to give the 1-loop
induced RG-invariant as it is calculated from the RG-equation in the
$O(N)$-models.  From the 1-loop invariant charge $$ \eqalign{
&Q(\hbox{${p^2\over \k^2}$},\hbox{${m^2 \over p^2}$} , \l)\cr & = \l +
\hbox{${1\over 16 \pi ^2} { N +8 \over 6}$} \l^2 \biggl(\, \sqrt { 1-
\hbox{${3 m^2 \over p^2}$} } \bigl( \ln \bigl( \sqrt{1- \hbox{${3 m^2 \over
p^2}$}} + 1 \bigr) - \ln \bigl( \sqrt{1- \hbox{${3 m^2 \over p^2}$}} - 1
\bigr) \bigr)\cr &\phantom{ = \l - \hbox{${1\over 16 \pi ^2} {3 \over 2}$}
\l^2}- \sqrt { 1- \hbox{${3 m^2 \over \k^2}$} } \bigl( \ln \bigl( \sqrt{1-
\hbox{${3 m^2 \over \k^2}$}} + 1 \bigr) - \ln \bigl( \sqrt{1- \hbox{${3 m^2
\over \k^2}$}} - 1 \bigr) \bigr)\biggr) + O(\l^3)\cr &\equiv \l - \l^2
\bigl(Q^{\scriptscriptstyle (1)} ( \frac {m^2}{p^2} ) - Q^{\scriptscriptstyle
(1)}( \frac {m^2}{\k^2} )\bigr) +O(\l^3),} \EQN{\loop1} $$ the RG-function $
\tilde \b _\l ^ {\scriptscriptstyle (1)} $ is calculated to be $$ \tilde \b
_\l^{\scriptscriptstyle (1)} ( \frac {m^2}{\k^2} ) = \hbox{${1\over 16 \pi ^2}
{N+ 8 \over 6}$} \l^2 \biggl( {3 m^2 \over \k^2 } { 1 \over {\sqrt { 1-
\hbox{${3 m^2 \over \k^2}$} }}} \ln { \sqrt{1- \hbox{${3 m^2 \over \k^2}$}} +
1 \over \sqrt{1- \hbox{${ 3 m^2 \over \k^2}$} } - 1} + 2 \biggr) \EQN{\BF1} $$
In the limit $ \k^2 \to -\infty$ the RG-function $ \tilde \b _\l ( \frac
{m^2}{\k^2} ) $ becomes $\k$-independent and coincides with the CS-function
$\b _\l^{\scriptscriptstyle (1)} $ and the one of the corresponding massless
models: $$ \b_\l ^{\scriptscriptstyle (1)} = \lim _{ \scriptscriptstyle \k^2
\to - \infty} \tilde \b _\l ^{\scriptscriptstyle (1)} ( \frac {m^2}{\k^2} ) =
\frac 1{16\pi^2} \frac {N+8}3 \l ^2, \EQN{\CS1} $$ In the limit $ \k^2 \to 0 $
the RG-function vanishes: $$ \lim_{\k^2 \to 0} \tilde \b _\l
^{\scriptscriptstyle (1)} ( \frac {m^2}{\k^2} ) = 0 \EQN{\A14} $$ With the
mass-dependent 1-loop RG-function $ \tilde \b _\l ^ {\scriptscriptstyle (1)} $
\queq{\BF1} one is able to solve the RG-equation \queq\RGQ \ analytically in
this approximation: $$\eqalign{ \widebar Q_1(\hbox{${p^2\over
\k^2}$},\hbox{${m^2 \over p^2}$} , \l) & = { \l \over 1 + \l(
Q^{\scriptscriptstyle (1)} ( \frac {m^2}{p^2} ) - Q^{\scriptscriptstyle (1)}(
\frac {m^2}{\k^2} ) ) }\cr & = \sum _{i=0}^\infty \l^{i+1} (
Q^{\scriptscriptstyle (1)} ( \frac {m^2}{\k^2} ) - Q^{\scriptscriptstyle (1)}(
\frac {m^2}{p^2} ) 	 )^i } \EQN{\QQ1} $$

$\widebar Q_1(\hbox{${p^2\over \k^2}$},\hbox{${m^2 \over p^2}$} , \l))$ is the
1-loop induced invariant charge; it continues in terms of 1-loop Green
function $ Q^{\scriptscriptstyle (1)} ( \frac {m^2}{p^2} ) -
Q^{\scriptscriptstyle (1)}( \frac {m^2}{\k^2} ) $ \queq{\loop1} the
renormalization group invariance to all orders.  Because we restrict ourselves
to the region where perturbation theory can be applied we understand this
non-perturbative solution always expanded as a power series in $\l$,
i.e.~treat it perturbatively.

The same form \queq{\QQ1} can be deduced for the spontaneously broken model
\queq{\F6sc}, where $Q^{\scriptscriptstyle (1)} ( \frac {m^2}{p^2} ) $ is a
different, more complicated function due to the appearance of further finite
diagrams.  In the asymptotic limit, $|\k^2| \gg m^2 $ and $|p^2| \gg m^2 $,
its invariant charge coincides with the one of the $O(2) $-model and its $\b
$-function $\tilde \b ^{\scriptscriptstyle (1)} _\l $ is the same as the one
of the CS-equation and the purely massless model \queq{\CS1}.  This is a
general feature of the asymptotic limit in these models, and -- as we will
point out -- it can be derived as a consequence of the fact, that a
CS-equation of the same form as the RG-equation exists.

Concluding the discussion of the 1-loop invariant charge we calculate the
action of the CS-operator on $\widebar Q_1(\frac {p^2}{\k^2} ,\frac {m^2}{p^2}
,\l) $ \queq{\QQ1}: $$ \bigl(\mdm + \kdk +\bl ^{\scriptscriptstyle (1)} \p _\l
\bigr) \widebar Q _1 (\frac {p^2}{\k^2} ,\frac {m^2}{p^2} ,\l) = \widebar Q_1^
2(\frac {p^2}{\k^2} ,\frac {m^2}{p^2} ,\l) \bigr( \bl ^{\scriptscriptstyle
(1)} - \tilde \bl ^{\scriptscriptstyle (1)} ( \frac {m^2}{p^2} ) \bigl),
\EQN{\CSQQ1} $$ where $\bl ^{\scriptscriptstyle (1)} $ is the Callan-Symanzik
function \queq{\CS1} and $\tilde \bl ^{\scriptscriptstyle (1)} $ is the
RG-function \queq{\BF1}.  As one easily verifies, the right-hand side is
well-defined: It is soft, i.e. it vanishes for asymptotic momenta, and it is a
RG-invariant in the same sense as $\widebar Q_1(\frac {p^2}{\k^2} ,\frac
{m^2}{p^2} ,\l)$.

\chap{The structure of the invariant charge} \sect{Renormalization group
invariants} In the last section we have calculated the 1-loop induced
invariant charge \queq{\QQ1} as the solution of the RG-equation with the
1-loop $\beta$-function.  In order to get deeper insight into the meaning of
it we will structure the higher order contributions according to their
invariance with respect to RG-transformations. For this purpose we consider
again the invariant charge as it is defined in \queq{\DEFQ} normalized
according to \queq{\NORMQ}.  (As in section~2 $p^2 $ and $\k^2 $ are always
taken in the Euclidean region for definiteness.)  $$ Q(\hbox{${p^2\over
\k^2}$},\hbox{${m^2 \over p^2}$} , \l)\Big|_{p^2=\k ^2} = \l \EQN{\NCQ} $$
Finite RG-transformations of the Green functions and especially of $Q$ can be
derived by a formal integration of the RG-equation \queq{\A4}, which expresses
the effect of infinitesimal RG-transformations in the differential form
\quref{\LowRG}.  But on a much more fundamental level invariance of the Green
functions under finite RG-transformations up to field redefinitions, the
anomalous dimensions, can be postulated directly as such \quref{\Bogo, \ZiRG}.
Due to its construction the invariant charge is an invariant under the
RG-transformations: If one has fixed $Q$ at a different point
$\k_{\scriptscriptstyle 1}^2$ calculating with a different coupling $\l
_{\scriptscriptstyle 1} $, $$ Q(\hbox{${p^2\over \k _{\scriptscriptstyle
1}^2}$},\hbox{${m^2 \over p^2}$} , \l _{\scriptscriptstyle
1})\Big|_{p^2=\k_{\scriptscriptstyle 1}^2} = \l _{\scriptscriptstyle 1},
\EQN{\NCQ'} $$ the RG-invariance requires that for all momenta the result has
to be the same: $$ Q(\hbox{${p^2\over \k^2}$},\hbox{${m^2 \over p^2}$} , \l)=
Q(\hbox{${p^2\over \k _{\scriptscriptstyle 1}^2}$},\hbox{${m^2 \over p^2}$} ,
\l _{\scriptscriptstyle 1}) \EQN{\RGDEF} $$ By means of the normalization
conditions we find $$ Q(\hbox{${\k_{\scriptscriptstyle 1} ^2\over
\k^2}$},\hbox{${m^2 \over \k_{\scriptscriptstyle 1}^2}$} , \l) =
\l_{\scriptscriptstyle 1} \EQN{\LL'} $$ from which one can immediately derive
the multiplication law of the RG: $$ Q(\tau \te , u, \l) = Q(\tau, u, Q(\te ,
u \tau , \l)) \EQN{\RGASS} $$ where $$\tau = \frac
{p^2}{\k_{\scriptscriptstyle 1} ^2} , \quad \te = \frac
{\k_{\scriptscriptstyle 1}^2}{\k^2} , \quad u = \frac {m^2}{p^2}.  $$ From now
on we restrict ourselves again to perturbation theory, where the invariant
charge is calculated in powers of the coupling $\l$ $$ Q(\tau , u, \l) =
\sum_{i=0}^\infty \l^{i+1} f_i(\tau , u) \EQN{\QPER} $$ where according to
\queq{\NCQ} $$ f_o (\tau , u) =1 \qquad \hbox{and} \qquad f_i(1 , u \tau) = 0
, \, i\ge 1 \EQN{\NCf} $$ Inserting the perturbatively calculated $Q(\tau , u,
\l) $ into the equation \queq{\RGASS} one gets $$ \sum_{i=0}^\infty \l^{i+1}
f_i(\tau \te, u) = \sum_{i=0}^\infty \Bigl( \sum_{j=0}^\infty \l^{j+1} f_j(\te
, u \tau) \Bigr)^{i+1} f_i(\tau , u) \EQN{\RGASSf} $$ Perturbatively we are
able to solve the equation recursively by comparing the same powers in the
coupling $\l$: $$\eqalign{ \l ^1:&\cr \l ^2:&\cr \l ^3:&\cr \l ^4:&\cr & \cr }
\qquad \eqalign { 1 & = 1 \cr f_1 (\tau \te, u) & = f_1 ( \te, u \tau ) + f_1
(\tau , u) \cr f_2 (\tau \te, u) & = f_2 ( \te, u \tau ) + f_2 (\tau , u) + 2
f_1 ( \te, u \tau ) f_1 (\tau , u) \cr f_3 (\tau \te, u) & = f_3 ( \te, u \tau
) + f_3 (\tau , u) + 3 f_2 (\tau , u) f_1 (\te, u \tau )\cr & + f_1 (\tau , u)
\bigl( f_1^2 ( \te, u \tau ) + 2 f_2 ( \te, u \tau)\bigr) \cr} \EQN{\REC} $$
The general expression in order k takes the form $$ \eqalign{ f_k (\tau \te,
u) & = f_k ( \te, u \tau ) + f_k (\tau , u) \cr & + \!\sum_{m=1}^{k-1}\!\! f_m
(\tau , u)\!  \!  \sum_{\(a_i\)_{m+1}} \! \!  (m\! +\! 1;a_1,...a_n,0)
f_o^{a_1} (\te, u \tau ) f_1^{a_2} (\te, u \tau ) ...  f_{\scriptscriptstyle
k-1}^{a_k} (\te, u \tau ) \cr} \EQN{\RECK} $$ where $$ (m+1;a_1,...a_l) = {
(m+1)!\over a_1!...a_l!} $$ and $\sum\limits _{\(a_i\)_{m+1}} $ is a sum over
all integer $ a_i \ge 0 $ with $a_1 + ... + a_k = m+1$ \lb \phantom{and
$\sum\limits_{\(a_i\)_{m+1}} $ a sum over all integer $ a_i \ge 0 $} and $a_1
+ 2a_2 + ... + k a_k = k+1.$

In this paper we will only evaluate the four lowest orders of perturbation
theory. But with the help of \queq{\RECK} it is a straightforward calculation
to show that the RG-solution \queq{\QQ1} fulfills the equations of finite
RG-transformations, too, where each order induces the next one necessarily.

The crucial equation appearing in the successive solution of the system
\queq{\REC} is the functional equation $$ f_k (\tau \te, u) = f_k ( \te, u
\tau ) + f_k (\tau , u) \EQN{\RGINV} $$ It is the starting equation of the
RG-invariant functions. A solution of it can be added in every order being
determined by true n-loop contributions.  In the massive case $(u \ne 0)$ the
general solution of \queq{\RGINV} is given by $$ f_k (\tau , u) = g _k(\tau u)
- g _k (u) \EQN{\RFS} $$ with $g_k (y) $ an arbitrary function of one
argument.  In the massless case $ (u=0 ) $ \queq{\RGINV} is much more
restrictive and the unique solution is the logarithm: $$ f_k (\t \te ) =
f_k(\t ) + f_k(\te ) \quad \Longleftrightarrow \quad f_k (\t) = \a _k \ln \t
\EQN{\RFSM} $$

In order to prove that \queq{\RFS} is the general solution of the functional
equation \queq{\RGINV} we rewrite it by a change of variables into $ (y = \t
\te , y' = \t)$ $$ f (y,u)-f (y',u)= f(\frac y{y'} ,uy') \eqnapp\RGINV ' $$ We
are now able to use the techniques of solving functional equations:
Differentiating \queq{\RGINV ' } with respect to $y, y'$ and $ u $ one gets
the following equations $(w = \frac y{y'} , v = u y')$: $$\eqalign{ y \p _y
f(y,u) & \, = \frac y{y'} \p _w f(w,v)\cr - y' \p _{y'} f(y',u) & \, = -\frac
y{y'} \p _w f(w,v)+ uy' \p_v f(w,v)\cr u \p _u f(y,u) -u \p _u f(y',u) & \, =
uy' \p _v f(w,v)\cr } \EQN{\RFDIFF} $$ Combining the equations in a way that
the explicit dependence on $w$ and $v$ cancels we remain with $$ y \p _y
f(y,u) -u \p _u f(y,u) = y' \p _{y'} f(y ' ,u) -u \p _u f(y',u) \EQN{\RFDIFF2}
$$ This is a partial differential equation with the left hand side independent
of $y'$ and the right-hand side independent of $y$ and consequently: $$ y \p
_y f(y,u) -u \p _u f(y,u) = F(u) \EQN{\RFDIFF3} $$ with the solution $$f(y,u)
= g(yu) + \tilde g (u) ,\quad \hbox{where} \quad u\p_u \tilde g(u) = F(u)
\EQN\RFGS $$ $ \tilde g (u) $ is fixed by inserting \queq\RFGS \ into
\queq{\RGINV '}: $$ \tilde g(uy) = - g(uy) \EQN\RFSS $$ and the general
solution of the functional equation \queq\RGINV \ is given by \queq{\RFS}.

In agreement with the explicit calculation of section 1 for the $O(N)$-models
we have in 1-loop (cf.~\queq{\loop1} with $ Q^{\scriptscriptstyle (1)} \equiv
g_1$ ) $$ Q^{\scriptscriptstyle (1)}(\frac {p^2}{\k^2} ,\frac {m^2}{p^2} , \l)
= \l^2 \Bigl(g_1(\frac {m^2}{\k^2} ) - g_1(\frac {m^2}{p^2} )\Bigr)
\EQN{\RF1l} $$ and for the massless case $$ Q^{\scriptscriptstyle
(1)}_{\scriptscriptstyle \infty} (\frac {p^2}{\k^2} ,\l) = \l^2 \a _1 \ln
\frac {p^2}{\k^2} \quad \hbox{with} \quad \a_1 = \frac 1{16\pi^2} \frac {N+8}6
\EQN{\RF1lm} $$ The solution of the next order works straightforwardly with
the result $$ f_2(\t ,u) = g_2 (\t u) - g_2(u) + (g_1(\t u ) - g_1 (u))^2
\EQN{\RF2l} $$ $ g_2 (\t u) - g_2(u)$ is a new true 2-loop function, starting
a new RG-invariant, whereas $(g_1 (\t u) - g_1(u))^2 $ continues the 1-loop
function to the next order to preserve the all order RG-invariance. It is
uniquely determined up to the addition of the RG-invariant $ g_1^2 (\t u) -
g_1^2(u)$.  The ordering we use here is guided to achieve in the RG-equation
minimal $\b$-functions, in a sense we will specify later (cf.~(3.26))

In a similar way one can solve the recursion formula \queq{\RECK} to calculate
the 3- and 4-loop order contribution: $$\eqalign{ f_3(\t,u) = & g_3 (\t u) -
g_3 (u) \cr + & \frac 52 (g_2 (\t u) - g_2(u))( g_1 (\t u) - g_1(u)) + \frac
12 (g_2 (\t u) g_1 ( u) - g_2 ( u) g_1 (\t u))\cr + & ( g_1 (\t u) - g_1(u))^3
\cr f_4(\t,u) = & g_4 (\t u) - g_4 (u) \cr + & 3 (g_3 (\t u) - g_3(u))( g_1
(\t u) - g_1(u)) + (g_3 (\t u) g_1 ( u) - g_3 ( u) g_1 (\t u))\cr + & \frac 32
( g_2 (\t u) - g_2 (u))^2 + \frac {13}3 (g_2 (\t u) - g_2(u))( g_1 (\t u) -
g_1(u))^2 \cr + & \bigl(\frac 53 g_1(\t u) -\frac 43 g_1( u )\bigr) \bigl(g_2
(\t u) g_1 ( u) - g_2 ( u) g_1 (\t u)\bigr)\cr + & ( g_1 (\t u) - g_1(u))^4 }
\EQN{\RF34l} $$ Before we discuss the structure of the RG-solution in more
detail we want to compare the result with the purely massless case in the same
model. One has $$ \eqalign{ f_1(\t) & = \a_1 \ln \t \cr f_2(\t) & = \a_2 \ln
\t +\a_1^2 \ln ^2 \t \cr f_3(\t) & = \a_3 \ln \t +\frac 52 \a_2 \a_1 \ln ^2 \t
+\a_1^3 \ln ^3 \t \cr f_4(\t) & = \a_4 \ln \t + ( 3 \a_3\a_1 + \frac 32 \a_2
^2 ) \ln ^2 \t + \frac {13}3 \a_2 \a_1^2 \ln ^3 \t + \a_1^4 \ln ^4 \t \cr}
\EQN{\RFM} $$ From here the structure is obvious: In every order there starts
a new RG-invariant, i.e.~a solution of \queq{\RGINV}: In the massless case it
is $\ln \t $ times an arbitrary coefficient which has to be determined by an
explicit evaluation of diagrams, in the massive case it is a new function of
the form $ g_n(\t u)- g_n (u)$. Besides this new invariant there appear
strings of lower order induced functions, which are series in $\ln \t $ in the
massless case, whereas in the massive case those are combinations of the
functions $g_i(u)$ and $g_k(u)$, which start to be interwoven with each other
from 3-loop onwards. Especially there appear also antisymmetric combinations
of the lower order functions, e.g.\ $g_2 (\t u) g_1 ( u) - g_2 ( u) g_1 (\t u)
$ in 3-loop order. The 1-loop induced solution of the RG-equation we have
calculated in \queq{\QQ1} is the RG-invariant induced string of the lowest
order. It has to appear in every order in addition to the higher order
contributions as it is required by RG-invariance. Especially one can verify
that it solves itself the functional equation \queq{\RECK}, if one puts all
higher order invariants equal to zero, i.e.\ $ g_i(u) = 0 $ for $ i \ge 2 $.

 We want to mention once again that the ordering of the lower loop induced
contributions is unique up to the addition of new invariants of the form
$g_{i_1}(\t u)\cdot...\cdot g_{i_n}(\t u) - g_{i_1}(u)\cdot...\cdot g_{i_n}(
u) $. As for $f_2(\t ,u)$ \queq{\RF2l} the structure we have imposed is again
due to the minimality of the $\b $-functions: Because $ Q(\t , u , \l) $ as
given in \queq{\RF1l,\shorttag\RF2l,\shorttag\RF34l} is constructed as an
invariant under finite RG-transformation, it is obvious, that it fulfills the
RG-equation, the infinitesimal form of \queq{\RGASS}: $$ \R Q = (\kdk + \tilde
\bl \p _\l ) Q = 0 \EQN{\RGQG} $$ Therefore we are able to calculate the $\b
$-functions $\tilde \bl $ up to four loops in terms of the RG-function $g_i
(y)$.  Applying \queq{\RGQG} on the 1-loop expression \queq{\RF1l}, one gets:
$$ - 2 \t u g_ 1 ' (\t u) + \tilde \b_\l ^{\scriptscriptstyle (1)} = 0
\Longrightarrow \tilde \b_\l ^{\scriptscriptstyle (1)} (\t u) = 2 \t u g_1
'(\t u ) \EQN{\BR1} $$ In the same way all the other $\b $-functions are
calculated: $$\eqaligntag{ \tilde \b ^{\scriptscriptstyle (2)}_\l (u \t ) = &
\l^3 2u \t g_2 '(u \t ) & \EQADV\RGBn\SUBEQBEGIN{\RGB2n} \cr \tilde \beta _\l
^{\scriptscriptstyle (3) } (u \t) = & \l ^4 \Bigl( 2 u \t g_3' (u\t) + \frac
12 g_1 (u\t ) 2u\t g_2'(u\t)- \frac 12 g_2(u\t ) 2 u \t g'_1(u\t) \Bigr) &
\SUBEQ{\RGB3n} \cr \tilde \b_\l^{\scriptscriptstyle (4)} ( \t u) = & \l ^5
\Bigl( 2 \t u g_4 ' (\t u ) + 2 \t u g_3 ' (\t u) g_1 (\t u ) - 2 \t u g_1 '
(\t u ) g_3 (\t u) \cr + & \frac 13 g_1 (\t u) \bigl( 2 \t u g_2 ' (\t u) g_1
(\t u) - 2 \t u g_1 ' (\t u) g_2 (\t u) \bigr) \Bigr) & \SUBEQ{\RGB4n} \cr }
$$ The ordering of the RG-invariants is due to the principle, that one has
avoided to introduce lower order RG-functions in the $\b $-functions, except
for the anti-symmetric combinations which are unavoidable and vanish if $g_i
(y) = c_i g_1 (y)$.  To make this statement of the minimality of $\b
$-functions clear, we add to the 2-loop solution \queq{\RF2l} an invariant of
the form $ g_1^2 (u\t) - g_1 ^2(u ) $ with an arbitrary coefficient $r$ $$
f_2(\t ,u) = g_2 (\t u) - g_2(u) + r (g_1^2 (u\t) - g_1 ^2(u ) ) + (g_1(\t u )
- g_1 (u))^2 \eqnapp{\RF2l}' $$ The 2-loop $\b $-function is then calculated
to be $$ \tilde \b ^{\scriptscriptstyle (2)}_\l (u \t ) = \l^3 \bigl( 2u \t
g_2 '(u \t ) + 2 r \tilde \b ^{\scriptscriptstyle (1)}_\l (u \t ) g_1( u\t )
\bigr) \eqnapp{\RGBn\RGB2n}' $$ i.e.~it contains a contribution appearing with
the 1-loop $\b$-function.  If one solves the RG-equation in the approximation
of the 1-loop $\b $-function, as we did in section 2, it is just assumed that
$\tilde \b ^{\scriptscriptstyle (2)}_\l $ is independent of $ g_1 (u \t) $ and
can be therefore neglected in some approximation. But, in section 6 we will
show, that through the CS-equation there arise exactly such terms in the
spontaneously broken model in the presence of fermions.

Furthermore, as it can be noticed from the 3-loop function, the form required
by the minimality of the RG-functions is at most compatible with the massless
theory: If all functions $g_i (y) $ tend to logarithms in the asymptotic
region: $$ g_i(y) \sim \ln (-y) \qquad \hbox{if} \quad y \to 0 \EQN{\HEPn} $$
the antisymmetric combinations vanish and the massless limit \queq{\RFM}
results to all orders. For the models introduced in section 2 the RG-functions
behave in fact as given in \queq{\HEPn}. But this result is not a consequence
of RG-invariance, but can be only derived by using the CS-equation.

\sect{The invariant charge at arbitrary momenta} In the defining equation of
the invariant charge \queq{\DEFQ} we have taken the momenta symmetrically, and
the Lorentz invariant combinations of external momenta are given therefore in
expressions of one momentum parameter $p^2 $.  In order to derive the string
structure of the invariant charge, this a unnecessary restriction. In general
the invariant charge of the scalar models depends of six independent Lorentz
invariants $$ Q = Q( \frac {p_1 ^2}{\k^2} ,\frac { m^2}{p_1 ^2} , P_{12} ,
P_{13} , P_{23}, P _{22} ,P _{33} ) \quad \hbox{with} \quad P_{ij} (\{p_k\})=
\frac {p_i \cdot p_j}{p_1 ^2} \EQN{\QGEN} $$ At the symmetric normalization
point $\k_{\scriptscriptstyle sym} $ (cf.~\queq{\A2}) the Lorentz invariants
are numbers $$ P_{ij} (\k _{\scriptscriptstyle sym}) \equiv K_{ij} = - \frac
13 \quad \hbox{if} \,\,\, i\ne j \quad \hbox{and} \quad P_{ii}
(\k_{\scriptscriptstyle sym} ) \equiv K_{ii} = 1 \EQN{\PIJsym} $$ Moreover
also the restriction to one mass parameter has been a simplification in order
to make the results more transparent, but is completely unnecessary.  A
further mass term can be consistently introduced for example in the
spontaneously broken model \queq{\F6sc}, if one breaks the symmetry explicitly
by a soft mass term for the Goldstone boson.  Fixing all the masses at the
pole, RG-invariance is defined as in \queq{\RGASS}: $$ Q(\tau \te , u, P_{ij}
, \a_l , \l) = Q(\tau, u,P_{ij}, \a_l, Q(\te , u \tau , K_{ij} , \a _l ,\l))
\EQN{\RGASSgen} $$ where now $ \tau = \frac {p_1 ^2}{\k^2} $, $u = \frac {m
^2}{p_1 ^2} $ and $\a _l $ denotes mass ratios $ { m_l ^2 \over m^2 } $. In
formula \queq{\RGASSgen} we have implicitly restricted the momenta to be in
the same sheet as the normalization point, i.e. they are Euclidean ones,
because we have assumed that there appears only one function Q.  Taking again
a perturbative power series expansion for the invariant charge the lowest
order equation, the general starting equation of RG-invariants, has the same
form as above \queq{\RGINV} except for the dependence on further parameters.
$$ f_k (\tau \te, u , P _{ij} , \a _l ) = f_k ( \te, u \tau , K_{ij} , \a _l )
+ f_k (\tau , u , P_{ij} , \a _l) \EQN{\RGINVgen} $$ If all momenta are taken
in the Euclidean region, as it is for the symmetric point, and the invariant
charge is real, it can be solved as above with the result: $$
Q^{\scriptscriptstyle (1)}(\tau , u, P_{ij} , \a_l , \l)= \l ^2 \bigl( g_1 (
\tau u, K_{ij} , \a_l ) - g_1 ( u, P_{ij} , \a_l ) \bigr) \EQN{\Q1lgen} $$
Having solved \queq{\RGINVgen} the solution of the respective 2-loop equation
\queq{\RF2l} can be calculated straightforward $$\eqalign{
Q^{\scriptscriptstyle (2) }(\tau , u, P_{ij} , \a_l , \l)= \l ^3 \bigl( & g_2
( \tau u, K_{ij} , \a_l ) - g_2 ( u, P_{ij} , \a_l ) \cr + & ( g_1 ( \tau u,
K_{ij} , \a_l ) - g_1 ( u, P_{ij} , \a_l ) )^2 \bigr)} \EQN{\Q2lgen} $$
Further information on momentum and mass dependence could be drawn from
symmetry properties of the Feynman diagrams, but such considerations are
beyond the purpose of this paper.  Already in such simple theories with
spontaneous breaking of the symmetry, there appear in 2-loop order a lot of
1-loop induced counterterms.  Thereby the knowledge of the general structure
\queq{\Q2lgen} should be helpful at least for a check of the results and for
the correct adjustment of finite counterterms having e.g.~calculated in a
scheme with asymptotic normalization conditions as it is the MS-scheme
\quref{\KrAs}.

\sect{The distribution of RG-invariants to diagrams} After having ordered the
n-loop terms according to their properties under RG-transformations the
question arises, if the lower loop induced contributions can be assigned to
certain Feynman diagrams. As we will point out, this structure can not be
associated to individual diagrams.  Especially the 1-loop induced invariant
charge $\widebar Q_1 $ \queq{\QQ1} is not just related to the sum of bubble
diagrams as it is often argued.  The situation seems to be rather analogous to
gauge invariance which is also not realized diagram by diagram but as a rule
only for Green functions.

In order to settle this issue we consider the 2-loop diagrams of the $ \phi ^4
$-theory. According to our construction of the invariant charge \queq{\DEFQ}
we have to add the self energy contribution $ \p _{p^2} \G _2 (p^2) $ to the
4-point function $ \G _4 $: $$ Q ^{\scriptscriptstyle (2)} (\frac {p^2}{\k ^2}
, \frac {m^2}{p ^2} , \l) = - \G _4 ^{\scriptscriptstyle (2)} (\frac {p^2}{\k
^2} , \frac {m^2}{p ^2} , \l) - 2 \l \p _{p^2} \G _2 ^{\scriptscriptstyle (2)}
(p^2,\frac {p^2}{\k ^2} , \frac {m^2}{p ^2} , \l) \EQN{\DEFQ2l} $$
RG-invariance of the invariant charge states that $ Q ^{\scriptscriptstyle
(2)} (\frac {p^2}{\k ^2} , \frac {m^2}{p ^2} , \l) $ has to be of the form
\queq{\RF2l} $$ Q ^{\scriptscriptstyle (2)} (\frac {p^2}{\k ^2} , \frac
{m^2}{p ^2} , \l) = Q ^{\scriptscriptstyle (2)} (\frac {m^2}{\k ^2} ) - Q
^{\scriptscriptstyle (2)} (\frac {m^2}{p ^2} ) + \Bigl( Q ^{\scriptscriptstyle
(1)} (\frac {m^2}{\k ^2} ) - Q ^{\scriptscriptstyle (1)} (\frac {m^2}{p ^2} )
\Bigr)^2 \eqnapp{\RF2l}' $$ $ Q ^{\scriptscriptstyle (1)} (\frac {m^2}{\k ^2}
) - Q ^{\scriptscriptstyle (1)} (\frac {m^2}{p ^2} ) $ is the 1-loop Green
function calculated in \queq{\loop1}, and $ Q^ {\scriptscriptstyle (2)} (y)
\equiv g_2(y)$, the genuine 2-loop function.

\line{} \PICT\frg {7cm} The 2-loop diagrams of the 4-point vertex in the
$\phi^4$-theory.

Calculating the bubble diagram fig.~1a in the BPHZL-scheme one gets: $$
\G^{\scriptscriptstyle {\rm (1a)}} _4 = - \frac 13 \l^3 \Bigl(
Q^{\scriptscriptstyle (1)} (\frac {m^2}{p ^2} ) -Q^{\scriptscriptstyle (1)}
(\frac {m^2}{p ^2} ) |_{p^2 = 0} \Bigr) ^2 \EQN{\DIA} $$ The subtraction at
$p^2 = 0 $ is due to the scheme we use.  From the counterterm inserted the
1-loop diagram (fig.~1d) one gets $$ \G^{\scriptscriptstyle {\rm (1d)}} _4 = 2
\l^3 \Bigl( Q^{\scriptscriptstyle (1)} ( \frac {m^2}{p ^2} )
-Q^{\scriptscriptstyle (1)} (\frac {m^2}{p ^2} ) |_{p^2 = 0} \Bigr) \Bigl(
Q^{\scriptscriptstyle (1)} (\frac {m^2}{\k ^2} ) -Q^{\scriptscriptstyle (1)}
(\frac {m^2}{\k ^2} ) |_{\k^2 = 0} \Bigr) \EQN{\DID} $$ The counterterm
diagram fig.~1e is independent of $p^2$ and in the BPHZL-scheme the diagram of
fig.~1c is zero. In order to obtain in 2-loop order the structure predicted by
the RG-invariance \queq{\RF2l '} the diagram of fig.~1b has to contain a
momentum dependent square of $ Q^{\scriptscriptstyle (1)} (\frac {m^2}{p ^2} )
$, $$ \G^{\scriptscriptstyle {\rm (1b)}} _4 = - \frac 23 \l^3 \Bigl(
Q^{\scriptscriptstyle (1)} ( \frac {m^2}{p ^2} ) \Bigr)^2 + \tilde
Q^{\scriptscriptstyle (2)} (\frac {m^2}{p ^2} ) + \hbox{constant} \EQN{\DIB}
$$ with $\tilde Q^{\scriptscriptstyle (2)} (y) \sim \ln y $ if $ y \to 0 $.
In \quref{\BePu} the diagram fig.\ 1b is calculated with the momenta taken on
mass-shell and the result confirms \queq{\DIB}. \lb In the massless theory,
where all momentum dependence is logarithmic at the symmetric point and the
structure is determined by \queq{\RFM}, one has in the BPHZL-scheme ($M^2$ is
the auxiliary mass) (cf.~for example \quref{\IZU}): $$\eqalign{
\G^{\scriptscriptstyle {\rm (1a)}} _4 & = -\frac 34 \Bigl( \frac 1{16\pi^2}
\Bigr)^2 \l^3 \Bigl( \ln | \frac {4 p^2}{3 M^2} | -2 \Bigr)^2 \cr
\G^{\scriptscriptstyle {\rm (1b)}} _4 & = \Bigl( \frac 1{16\pi^2} \Bigr)^2
\Bigl( -\frac 32 \ln^2 | \frac {4 p^2}{3 M^2} | + 9 \ln | \frac {4 p^2}{3 M^2}
| + C \Bigr) \cr \G^{\scriptscriptstyle {\rm (1c)}} _4 & = 0 \cr
\G^{\scriptscriptstyle {\rm (1d)}} _4 & = \frac 92 \Bigl( \frac 1{16\pi^2}
\Bigr)^2 \Bigl( \ln | \frac {4 p^2}{3 M^2} | -2 \Bigr) \Bigl( \ln | \frac {4
\k^2}{3 M^2} | -2 \Bigr) \cr \G^{\scriptscriptstyle {\rm (1e)}} _4 & = \Bigl(
\frac 1{16\pi^2} \Bigr)^2 \Bigl(- \frac 94 \ln ^ 2 | \frac {4 \k^2}{3 M^2} | +
6 \ln | \frac {4 \k^2}{3 M^2} | - \tilde C \Bigr) \cr }\EQN{\DIM} $$ The
2-point function is given by $$ \G_2 (p^2,\k^2, \l) = - \frac 1{12} \Bigl(
\frac 1{16\pi^2} \Bigr)^2 \Bigl(p^2 \ln \frac {p^2}{\k^2} - p^2\Bigr)
\EQN{\arsch} $$ Only the sum of all the diagrams has the desired form: $$ Q
^{\scriptscriptstyle (2)} (\frac {p^2}{\k ^2} , \frac {m^2}{p ^2} , \l) =
\Bigl( \frac 1{16 \pi ^2} \Bigr) ^2 \Bigl( \frac 94 \ln^2 \frac {p^2}{\k^2} -
\frac {17}6 \ln \frac {p^2}{\k^2} \Bigr) \EQN{\FINQ2} $$ It is independent of
the auxiliary mass $M^2$ and the square of the logarithm appears in
RG-invariant form, i.e.\ its coefficient is the 1-loop coefficient squared
(cf.~\queq{\RF1lm}).  The correct coefficient arises only when the diagrams
fig.~1a and 1b are summed up. The square of the logarithmic terms is in all
schemes uniquely associated with fig.~1a and 1b, whereas the single
logarithmic term originates from different diagrams in different schemes. The
appearance of the 1-loop induced contributions seems to us to be strongly
related to the assignment of sectors to diagrams as they have been introduced
e.g.~in the framework of dimensional regularization \quref{\BreiMai}. If such
a technique is helpful also for singling out finite contributions remains to
be clarified.

\chap{The CS-equation} In the last section we have shown, how the requirement
of RG-invariance structures the invariant charge.  To get further information
on the RG-functions $g_i (y) $ we have to use the CS-equation. In this section
we consider such theories, as they have been introduced in section 2, where
the CS-equation has the same form as the RG-equation: $$ \C Q = (\mdm + \kdk +
\bl \p _\l ) Q = Q _m \eqnapp{\CSQ}' $$ We apply the CS-operator order by
order on the invariant charge as it is given in
\queq{\RF1l,\shorttag\RF2l,\shorttag\RF34l} and calculate $\b_ \l $ in
expressions of the RG-functions $g_i(y)$. Thereby we will derive the
high-energy behavior we have mentioned in the last section \queq{\HEPn} and
mass independence of the $\tilde \bl $ and $\bl $ for asymptotic normalization
conditions.  Besides this we will not find any further restrictions on the
RG-functions, $g_i(y), i=1..4 $ and the CS-equation is completely consistent
with RG-invariance by itself.

Starting point is the 1-loop solution \queq{\RF1l} $$ Q^{\scriptscriptstyle
(1)}(\t,u , \l) = \l^2 \Bigl(g_1(u \t) - g_1(u )\Bigr) \EQN{\RF1lm} $$ with $u
= \frac {m^2}{p^2} $ and $\t =\frac {p^2}{\k^2} $ ($p^2$ and $\k^2 < 0 $ as
above).  Inserting \queq{\RF1lm} into the CS-equation gives $$ -2 u g_1 '(u)
\l^2 + \b _\l ^{\scriptscriptstyle (1)} = Q_m^{\scriptscriptstyle (1)} (u)
\EQN{\CSQ1} $$ The CS-equation has to be valid for all momenta $p^2$,
therefore especially for large ones, where the right-hand side will vanish
according to its construction, if one is not at an exceptional momentum. The
symmetric point is non-exceptional and one has: $$ \l^2 \lim_{u \to 0} 2 u g_1
' (u) = \b _\l ^{\scriptscriptstyle (1)} \equiv b_o \l ^2 \EQN{\BCS1} $$ From
here it follows that $ \b_\l ^{\scriptscriptstyle (1)} $ is independent of $
\kappa ^2 $, i.e.\ a constant, as it is well-known.  Furthermore by
integration one derives that $g_1 (u) $ has logarithmic behavior for
asymptotic momenta: $$ g_1(u) = \frac 12 b_o \ln(- u) + C_1 \quad \hbox{for}
\quad u \to 0 \EQN{\AB1} $$ Therefore the asymptotic behavior of the 1-loop
function $g_1 (y) $ is in fact a consequence of the existence of the
CS-equation.  Because RG-invariance has only determined the difference $
g_1(\t u) - g_1 (u )$ we are free to normalize $g_i(y) $ in such a way that
$C_1=0 $ for $y \to 0 $.  This normalization leads to some simplification
later on, especially it simplifies the transition to the massless theory.
 From \queq{\BCS1} and \queq{\BR1} it follows also the result we have mentioned
in section~2, that the $\b$-function of the RG-equation \queq{\BR1} is the
same as the CS-$\b $-function in the limit of an asymptotic normalization
point: $$ \lim _{\k^2\to -\infty} \tilde \b _\l ^{\scriptscriptstyle (1)} (
\frac {m^2}{\k^2} ) = \b _\l ^{\scriptscriptstyle (1)} \EQN{\ASBR} $$ In
agreement with our explicit calculation (cf.~\queq{\CSQQ1}) one finds for the
right-hand side, the soft mass insertion: $$ Q_m^{(1)} (u) = \b_\l
^{\scriptscriptstyle (1)} - \tilde \b_\l ^{\scriptscriptstyle (1)} (u)
\EQN{\QM1} $$ It is $\kappa $-independent as required by the consistency
equation of the RG- and the CS-equation: $$ \( \R ,\C \) Q = \R Q _m
\EQN{\CONS} $$ which means in 1-loop $$ \kdk \b^{\scriptscriptstyle (1)} _\l =
\kdk Q_m^{(1)} = 0 \EQN{\CONS1} $$

The 2-loop order works in the same way as the 1-loop order.  Taking the
$Q^{\scriptscriptstyle (2)} (\t ,u , \l )$ of \queq{\RF2l } one finds for the
CS-$\b $-function $$ \b^ {\scriptscriptstyle (2)}_\l = \l^3 \lim _{u \to 0} 2
u g_2 (u) \EQN{\BRC2} $$ Therefrom we deduce the same results as above: The
$\beta $-function of the CS-equation in 2-loop order is $\kappa $-independent
again, consequently in the limit $u \to 0$ the function $g_2 (u) $ has
logarithmic behavior. For an asymptotic normalization point the
$\beta$-function of the CS-equation and of the RG-equation \queq{\RGBn \RGB2n}
are equal.  $$ \beta _\l ^{\scriptscriptstyle (2) } = \lim _{
\scriptscriptstyle \k^2 \to -\infty} \tilde \beta _\l ^{\scriptscriptstyle (2)
} ( \frac{m^2}{\k^2} ) \equiv b_1 \l ^3 \quad \hbox{and} \quad \lim _{u\to 0}
g_2 (u) = \frac 12 b_1 \ln(- u) \EQN{\AS2} $$ where we have chosen the
arbitrary integration constant to be zero, using the same arguments as above
(cf.~\queq{\AB1}).  The right-hand side of the CS-equation can be calculated
in terms of the functions $g_i (y) $ and their derivatives again: $$
Q_m^{\scriptscriptstyle (2)} (\t,u) = \beta _\l ^{\scriptscriptstyle (2) } -
\tilde \beta _\l ^{\scriptscriptstyle (2) } (u) + 2 \bigl( \beta _\l
^{\scriptscriptstyle (1) } -\tilde \beta _\l ^{\scriptscriptstyle (1) } (u)
\bigr) \bigl(g_1 (\t u) - g_1(u)\bigr) \EQN{\QM2} $$ According to \queq{\AS2,
\shorttag\AB1,\shorttag \ASBR} in the limit of asymptotic $p^2 $, i.e. $u \to
0 , u\t = \frac {m^2}{\k^2} = $ const., $Q_m^{\scriptscriptstyle (2)} $ is
vanishing and therefore soft, as required.  All these findings are in complete
agreement with the consistency equation \queq{\CONS} of order 2-loop: $$ 0=
\kdk \beta _\l ^{\scriptscriptstyle (2) } = \kdk Q_m^{\scriptscriptstyle (2)}
(\t,u) + \tilde \beta _\l ^{\scriptscriptstyle (1) } (\t u ) \p _\l
Q_m^{\scriptscriptstyle (1)} (\t,u) \EQN{\CONS2} $$ No further restrictions on
$g_2(y) $ and $ g_1(y)$ are required.

Due to the anti-symmetric combinations we have found in the RG-solutions from
3-loop onwards \queq{\RF34l} the structure starts to become more complicated,
especially the $\beta $-function of the CS-equation starts to become $\kappa
$-dependent as expected.  Using the arguments as above the CS-function is
found to be $$ \beta _\l ^{\scriptscriptstyle (3) } (u\t) = \l^4 \Bigl( \lim
_{u \to 0} \bigl( 2u g_3 ' (u) - \frac 12 b_1 g_1 (u) + \frac 12 b_o g_2
(u)\bigr) +b_1 g_1 (\t u) - b_o g_2 (\t u ) \Bigr) \EQN{\CSB3} $$ Therefore
$\b_\l^{\scriptscriptstyle (3)} $ is $\kappa $-dependent, but the momentum
dependent terms have to tend to a constant in the asymptotic limit, $$ \lim
_{u \to 0} \bigl( 2u g_3 ' (u) - \frac 12 b_1 g_1 (u) + \frac 12 b_o g_2
(u)\bigr) \equiv b_2 \EQN{\ASYB3} $$ With the normalization of the functions $
g_1 (u) $ and $ g_2 (u) $ we have chosen in \queq{\AB1, \shorttag \AS2} ($C_1
= 0 $) one has: $$ \lim_{u\to o}(b_1 g_1 (u) - b_o g _2 (u) ) = 0 $$ and we
find the logarithmic behavior of $g_3 (u) $ $$\lim _{u \to 0 } g_3 (u) = \frac
12 b_2 \ln(-u) \EQN{\AS3} $$ For asymptotic $ \k ^2 $ the CS-function is
independent of $\k $ and agrees with the RG-function \queq{\RGBn \RGB3n} $$
\lim _{\scriptscriptstyle {\k^2 \to - \infty}} \beta _\l ^{\scriptscriptstyle
(3)} (\frac {m^2}{\k^2} )= b_2 \l^4 = \lim _{\scriptscriptstyle {\k^2 \to -
\infty}} \tilde \beta _\l ^{\scriptscriptstyle (3)} (\frac {m^2}{\k^2} )
\EQN{\ABRC3} $$ For finite $\kappa $ the $\k $-dependence of $\beta _\l
^{\scriptscriptstyle (3) } $ gives a measure, how far the 2-loop function $g_2
(u)$ and 1-loop function $g_1 (u) $ differ in the mass-dependent region.
Especially it is just the antisymmetric term in \queq{\RF34l}, which causes
the $\k$-dependence of $\b_\l^{\scriptscriptstyle (3)} $. E.g.~in the case, if
$g_2 (u) = \frac {b_1}{b_o} g_1(u) $ for all $u$, the antisymmetric
contribution as well as the $\k $-dependence in the $\b$-function of the
CS-equation would cancel.\lb Finally looking into the consistency equation
\queq{\CONS} of 3-loop order one verifies the following identities, which are
valid without any restrictions on $g_i(x)$: $$\eqalign{ & \kdk \b _\l
^{\scriptscriptstyle (3)} (\frac {m^2}{\k^2} )= \tilde \beta _\l
^{\scriptscriptstyle (2) } (\frac {m^2}{\k ^2} ) \beta _\l
^{\scriptscriptstyle (1) } - \tilde \beta _\l ^{\scriptscriptstyle (1) }
(\frac {m^2}{\k^2} ) \beta _\l ^{\scriptscriptstyle (2) } \cr \hbox{and} & \cr
& \Bigl(\bigl(\kdk + \tilde \b _\l ( \frac {m^2}{\k^2} ) \p_\l \bigr) Q_m
\Bigr) ^{(3)} = 0 } \EQN{\CONS3} $$

For the 4-loop order \queq{\RF34l} the calculations work in the same way as
for the 3-loop order and we only state the result: $$\eqalign{
\b_\l^{\scriptscriptstyle (4)} ( \t u) = \l^5 \Bigl( &\lim _{u \to 0} \bigl(2
u g_4 ' ( u )\bigr) \cr + & 2 \bigl(b_2 g_1 (\t u ) - b_o g_3 (\t u) \bigr) +
g_1 (\t u) \bigl(b_1 g_1 (\t u) - b_o g_2 (\t u)\bigr) \Bigr)} \EQN\RGB4b $$
Therefore, the RG-function $g_4 (u)$ starting the new RG-invariant of order
4-loop has logarithmic behavior for $u \to 0 $, i.e.~in the massless limit, $$
\lim _{u \to 0 } u g_4' ( u ) \equiv b_3 \;\Longrightarrow \; g_4(u) = \frac
12 b_3 \ln(- u) \quad \hbox{for} \quad u \to 0 \EQN\ASG4 $$ taking the
normalization for the integration constant as above \queq\AS2.  In the limit
of an asymptotic normalization point the $\b$-functions \queq{\RGBn\RGB4n} and
\queq\RGB4b \ agree and are given by the constant $b_3$.  All the results are
again in perfect agreement with the consistency equation \queq{\CONS} of
4-loop order: $$\eqalign{ & \kdk \b _\l ^{\scriptscriptstyle (4)} (\frac
{m^2}{\k^2} )= 2 \bigl( \tilde \beta _\l ^{\scriptscriptstyle (3) } (\frac
{m^2}{\k ^2} ) \beta _\l ^{\scriptscriptstyle (1) } - \tilde \beta _\l
^{\scriptscriptstyle (1) } (\frac {m^2}{\k^2} ) \beta _\l ^{\scriptscriptstyle
(3) } ( \frac {m^2}{\k^2} ) \bigr)\cr \hbox{and} \qquad & \cr &
\Bigl(\bigl(\kdk + \tilde \b _\l ( \frac {m^2}{\k^2} ) \p _\l \bigr) Q_m
\Bigr) ^{(4)} = 0 } \EQN{\CONS4} $$

In a purely massive theory, e.g.~massive $\phi^4 $-theory, where all functions
have to exist at $p^2 =0 $, too, one can simply derive that the RG-$\b
$-functions vanishes at $\k^2 = 0$. Therefore one has to test the CS-equation
at $p^2 = 0 $ with the explicit expressions for the $\b $-functions and the
soft insertion on the right-hand side using that the differential operator
$\mdm + \kdk $ commutes with the test at $p^2 =0$ $$ \lim _{\k^2 \to 0} \tilde
\b ^{\scriptscriptstyle (i)}_\l (\frac {m^2}{\k^2} ) = 0 \quad i=1,...,4
\qquad \hbox{if all fields are massive} \EQN{\RBK0} $$

Although we did not succeed to find an all order recursion formula we are
convinced that the same results will be achieved in any order of perturbation
theory, especially the structure of RG-strings and their associated
logarithmic high-energy behavior. At the same time we do not expect any
further restrictions on the functions $g_i(x) $ following from consistency of
the RG- and CS-equation. Therefore a RG-solution with an appropriate
high-energy behavior seems to be automatically consistent with the CS-equation
in the one coupling theory.

With these results it is obvious that the asymptotic limit goes smoothly into
the logarithmic structure of the massless theory \queq{\RFM} with $\alpha _i =
\frac 12 b _{i-1} $.  Having derived the string structure of the RG-invariant
solution and the logarithmic behavior of the $g_i(u)$ the use of improvement
for a theory with a positive $\b$-function as mentioned in the introduction
can be clarified: If one has fixed the coupling at a specific normalization
point $\k^2 _o = c_o m^2$ for example in the low energy region and takes the
limit to large $p^2$, all the momentum dependent terms tend to logarithms.  In
each order the highest power in the logarithms is given by the power of $ g_1
(\frac {m^2}{p^2} )$, all the other momentum dependence has lower powers in
the logarithms. Therefore as long as perturbation theory is valid, the 1-loop
induced invariant charge will dominate all the other contributions for large
$p^2 $.

\chap{Reparametrizations of the coupling} The structure of the invariant
charge we have found in \queq{\RF1l,\shorttag\RF2l,\shorttag\RF34l} is related
to the choice of normalization conditions \queq{\NCQ} identifying the
invariant charge with the coupling itself at a normalization point $p^2 = \k^2
$ to all orders.  Calculations carried out in schemes without specific
normalization conditions can change this structure due to the fact that one is
not able to find such a point $\k^2 $ to all orders. But in any allowed scheme
there has to be a point $p^2 = \k^2 $ -- mostly in the asymptotic region --
where the invariant charge is related to a power series in the coupling
\foot{$\rho _1$ can be always made to vanish by chosing $p^2 = c \k^2$, with
$c$ a number.}: $$ Q(\hbox{${p^2\over \k^2}$},\hbox{${m^2 \over p^2}$} ,
\l')\Big|_{p^2=\k ^2} = \l' + \rho _1 \l'^2 + \rho _2 \l'^3 + ...
\EQN{\NCQre} $$ The change in the structure of the invariant charge arising
from such normalization conditions can be taken into account by considering
reparametrizations of the coupling $\l$: $$ \l(\l') = \l' + \rho _1 \l'^2 +
\rho _2 \l'^3 + ...  \EQN{\REP} $$ For the invariant charge calculated with
the redefined coupling $$ Q(\hbox{${p^2\over \k^2}$},\hbox{${m^2 \over p^2}$}
, \l') = \sum _{k=0} ^\infty \l '^{k+1} f_k (\t , u, \rho_i) $$ one gets
immediately the following expression. $f_k(\t,u) $ denotes the functions of
the properly normalized invariant charge as given in
\queq{\RF1l,\shorttag\RF2l,\shorttag\RF34l}: $$\eqalign{ \l '^1:&\cr \l
'^2:&\cr \l '^3:&\cr \l '^4:&\cr \l'^5: & \cr & \cr } \quad \eqalign { f_o
(\tau, u, \rho _i) & = 1 \cr f_1 (\tau , u ,\rho_i) & = \rho_1 + f_1 (\tau ,
u) \cr f_2 (\tau , u ,\rho _i) & = \rho _2 + 2 \rho_1 f_1 (\tau , u) + f_2
(\tau , u)\cr f_3 (\tau , u ,\rho _i) & =\rho _3 + (\rho _1 ^2 + 2 \rho _2 )
f_1 (\tau , u) + 3 \rho _1 f_2 (\tau , u) + f_3 (\tau ,u) \cr f_4 (\tau , u
,\rho _i) & =\rho _4 + 2(\rho _2\rho _1 + 2 \rho _3 ) f_1 (\tau , u) + 3( \rho
_2 + \rho_1^2 ) f_2 (\tau , u) \cr & + 4 \rho_1 f_3 (\tau ,u) + f_4 (\tau , u)
\cr} \EQN{\QREP} $$ The leading contribution $(g_1 (u\t) -g_1(u)\bigl)^n$ in
the n-loop order is not affected by such a reparametrizations, whereas
asymptotically the coefficients of all the lower logarithms are shifted by
constants. The general structure remains: the starting of a new
renormalization group invariant and the necessary appearance of lower order
introduced functions.  The $\b $-functions of the RG-equation and of the
CS-equation are changed according to the well-known formula: $$ \b _{\l'}
(\frac {m^2}{\k^2} , \rho _i) = \Bigl( \frac {d\l}{d \l'}
\Bigr)^{\scriptscriptstyle -1} \b_{\l (\l ')} (\frac {m^2}{\k^2} ) \qquad
\hbox{and} \qquad \tilde\b _{\l'} ( \frac {m^2}{\k^2} ,\rho _i) = \Bigl( \frac
{d\l}{d \l'} \Bigr)^{\scriptscriptstyle -1} \tilde \b_{\l(\l ')} ( \frac
{m^2}{\k^2} ) \EQN\BETAREP $$ As it is well-known the two lowest orders of the
$ \beta $-functions are not changed by such a reparametrization but are given
with their $\kappa $-dependence quite generally by \queq{ \BR1 ,\shorttag
\RGBn \RGB2n} and \queq {\BCS1 ,\shorttag\BRC2}. The 3-loop order is changed
to $$ \tilde \b _{\l'}^{\scriptscriptstyle (3)} (\frac {m^2}{\k^2} , \rho_ i)
= \tilde \b_{\l'}^{\scriptscriptstyle (3)} (\frac {m^2}{\k^2} )+ \rho _1
\tilde \b_{\l'}^{\scriptscriptstyle (2)}(\frac {m^2}{\k^2} ) - \rho _2 \tilde
\b_{\l'}^{\scriptscriptstyle (1)} (\frac {m^2}{\k^2} )+ \rho _1 ^2 \tilde
\b_{\l'}^{\scriptscriptstyle (1)} (\frac {m^2}{\k^2} ) \EQN\BETAREP3 $$ with
the equivalent formula for the CS-$\beta$-function, too.  One should notice
that in a massive theory with general non-asymptotic normalization conditions
one cannot make the $\beta $-functions of the CS- or RG-equation vanish from
3-loop order onwards by a reparametrization as it is possible for the massless
theory, because the explicit $\k $-dependence will just be canceled if $ g_1
(y ) \sim g_2 (y) $. Such a cancellation would be intrinsic to a theory and
could not be forced from outside.

\chap{The asymptotic invariant charge of the scalar particle \nl in the
spontaneously broken $U(1)$--axial model} In this section we will apply the
methods developed above to the spontaneously broken $U(1)$-axial model with a
scalar/pseudoscalar doublet $A/B $ and one fermion $\psi $ \quref{\CalSym};
the classical action is given by $$\eqalign{ \Gacl = \int \Bigl( & \ha
\bigl(\p A \p A + \p B \p B \bigr) + i \bpsi \dslash \psi -
m_{\hbox{\eightpoint $f$}} \bpsi \psi - {m_{\hbox{\eightpoint $f$}} \over
m_{\hbox{\eightpoint $H$}}} \sqrt {\frac { \l}{3} } \bpsi (A + i \gamma _ 5 B
)\psi \cr & - \frac 12 m_{\hbox{\eightpoint $H$}}^2 A^2 - \frac 12
m_{\hbox{\eightpoint $H$}} \sqrt{ \frac \l 3 } A ( A^2 + B^ 2) - \frac \l{4!}
( A^2 + B^ 2)^2 \Bigr)} \EQN{\F6} $$ In contrast to the models considered in
section~4 the CS-equation and RG-equation do not have the same structure,
i.e.~do not contain the same hard differential operators in the physical
parametrization, where the physical masses are fixed at the pole of the
respective propagators.  But from the point of view of the RG-equation it is a
one-coupling theory as the models considered above. Therefore the whole
analysis of sect.~3 is valid for the invariant charge of the scalar coupling
especially the ordering according to RG-invariant functions can be applied as
it is. The crucial point is the action of the CS-equation on the RG-invariant
functions $g_i(y) $. In order to simplify the analysis and to get a first
impression of the consequences of the physical mass $\beta $-function in the
CS-equation we consider the invariant charge normalized at an asymptotic
normalization point and consequently at asymptotic momenta.  As a result we
derive that the RG-functions $g_i(y) $ contain well-defined powers of
logarithms in higher orders inducing at the same time a certain
$\k$-dependence into the $\beta $-functions, which does not vanish in the
asymptotic region.

Before turning to the calculations a few words about the normalization
conditions we impose are in order: Throughout the paper we have normalized the
masses on-shell. Especially in the spontaneously broken models considerations
concerning the RG-equation are often presented in the symmetric way specifying
the couplings and taking the shift parameter $v$ parametric in the
Ward-identity. If such considerations are carried out with mass-independent
$\b$-functions one calculates the symmetric limit.  Moreover there is no way
to reach the massive region by a renormalization group integration for such
parametrizations even in lowest order or in any approximation, because
typically terms as $\frac {\l v}{\k^2} \ln \frac {\l v }{\k^2} $ appear in the
$\beta$-functions.  This is not just a annoying technical problem, but it is
deeply connected with the fact, that the order by order finite RG-invariance
in the sense we have formulated in sect.~2 is lost.  In order to detect finite
RG-invariance in perturbation theory one has to expand the RG-functions
$g_i(y) $, i.e.~the Green functions, as a Taylor series in the coupling: $$
g_i(\frac {m^2}{p^2} + \rho \l \frac {m^2}{p^2} ) = g _i ( \frac {m^2}{p^2} )
+ \rho \l \frac {m^2}{p^2} g'_i(\frac {m^2}{p^2} ) + ...  \EQN{\tay} $$ Such
an expansion is only valid for a restricted range of momenta. Therefore
normalization conditions, which do not fix the pole of the propagator, seem to
us to make the perturbative expansion even worse than it is expected to be
anyway.

Apart from these theoretical aspects the $U(1) $-axial model can be considered
as a toy model of the matter sector of the standard model.  One reason for
carrying out such a RG-analysis is to get estimates of the higher orders, as
e.g.~of the normalization point dependence, if one knows the 1-loop order
completely. With this aim in mind one has to parametrize in a way as it is
done in realistic models. In the standard model the masses of the particles
are known to a high accuracy and are therefore taken as an experimental input
for all further calculations.  Only with such normalization conditions one can
expect to get useful results for the 2- or 3 loop order.

The Green functions of the model are constructed according to the Gell-Mann
Low formula and with a suitable renormalization prescription, which we do not
specify, because we just consider the finite Green functions.  The model is
defined by the requirement that the Green functions fulfill the Ward-identity
of the spontaneously broken $U(1) $-axial symmetry $$ \bW \G = 0 \quad
\hbox{with} \quad \bW = -i \int \bigl( ( A + v ) {\d \over \d B} - B {\d \over
\d A} - {i \over 2 } {\d \over \d \psi } \ga _5 \psi - {i \over 2 }\bpsi \ga
_5 {\d \over \d \bpsi } \bigr), \EQN{\F2} $$ and by normalization conditions
to fix the free parameters of the theory: $$\eqalign{ & \p _{p^2} \Gamma_{AA}
\big|_ {p^2 = \kappa ^2} = 1 \qquad \gamma^ \mu \p _{ p ^\mu} \Gamma _ {\psi
\bpsi} \big| _ { \pslash = \kappa} = 1 \qquad \Gamma_ A = 0 \cr & \Gamma _{AA}
\big| _{p^2 = m_{\hbox{\eightpoint $H$}}^2} = 0 \qquad \Gamma _ {\bpsi \psi}
\big|_ {\pslash = m_{\hbox{\eightpoint $f$}}} = 0 \cr & \G _{AAAA}\big|_
{\{p_i\} = \kappa _{\hbox{\SMALL {sym}}} } = - \lambda } \EQN\norm $$ With
these normalization conditions the Green functions are calculated
perturbatively in powers of the coupling $\sqrt \l$, the shift parameter is
determined by the Ward-identity: $$ \hat v \equiv \hat v (m_{\hbox{\eightpoint
$H$}},m_{\hbox{\eightpoint $f$}},\k , \l ) = \sqrt { \frac { 3}{ \l} }
m_{\hbox{\eightpoint $H$}} + O (\hbar ) \EQN\shift $$ In ref.~\quref{\CalSym }
we have shown that the CS-equation exists to all orders with a soft insertion
on the right-hand side: $$ \bigl( m \p_m + \b_{m_{\hbox{\eightpoint $f$}}}
m_{\hbox{\eightpoint $f$}} \p _ {m_{\hbox{\eightpoint $f$}}} + \b _\l \p _\l -
\ga _ B \N _ B - \ga_F \N _ F \bigr) \G = \hat v (1 + \rho ) \int \bigl( {\d
\over {\d A}} +\hat \a {\d \over {\d q}}\bigr) \G_{\big|_{q=0}} \EQN{\G46} $$
where $q$ is an external field coupled to the invariant of infrared dimension
2 and $$\eqalign{& m \p_m = m_{\hbox{\eightpoint $H$}} \p
_{m_{\hbox{\eightpoint $H$}}} + m_{\hbox{\eightpoint $f$}} \p
_{m_{\hbox{\eightpoint $f$}}} +\k \p _\k \cr & \N _B = \int \bigl( A{ \d \over
\d A} + B{ \d \over \d B}\bigr) \qquad \N _F = \int \bigl( \psi{\d \over \d
\psi } + \bpsi {\d \over \d \bpsi} \bigr)\cr & \hat v \rho = ( \b _ \l \p_\l +
\b_{\mf} \mf \p _ {\mf} + \ga_B) \hat v} \SUBEQNBEGIN{\G47} $$ The
renormalization group equation has due to the physical normalization condition
the simple form $$ \Bigl( \kappa \p _ \kappa + \tilde \b _ \l \p _ \l -\tilde
\ga _ B \N _ B - \tilde \ga _ F \N_ F \Bigr) \G = 0 \EQN{\G48} $$ with the
additional constraint: $$ \kappa \p _ \kappa \hat v + \tilde \b_ \l \p _\l
\hat v + \tilde \ga _ B \hat v = 0 \SUBEQNBEGIN{\G49} $$ The invariant charge
of the scalar field $A$ defined according to \queq{\DEFQ} with $\G _2 \equiv
\G _ {\scriptscriptstyle AA} $ and $ \G _4 \equiv \G_ {\scriptscriptstyle
AAAA} $ has the same properties as the invariant charge of the scalar models:
It is dimensionless $$ Q ( p^2 , m_{\hbox{\eightpoint $f$}} ,
m_{\hbox{\eightpoint $H$}} , \k , \l) = Q ( \frac {p^2}{\k^2} ,
\frac{m_{\hbox{\eightpoint $H$}} ^2}{p^2} , \frac {m_{\hbox{\eightpoint $f$}}
}{m_{\hbox{\eightpoint $H$}}} , \l) , \EQN{\Qdim} $$ it has well-defined
normalization properties $$ Q ( \frac {p^2}{\k^2} , \frac{m_{\hbox{\eightpoint
$H$}} ^2}{p^2} , \frac {m_{\hbox{\eightpoint $f$}} }{m_{\hbox{\eightpoint
$H$}}} , \l) \Big| _ {p^2 = \k^2} = \l \EQN{\Qnorm} $$ and furthermore it is a
RG-invariant, i.e.~it satisfies the homogeneous RG-equation and the
CS-equation without anomalous dimension.

A complete analysis of the 1-loop induced higher order contributions requires
the complete knowledge of the finite Green functions in 1-loop order.  All
these Green functions are calculated in the literature but for a first
approach we want to restrict ourselves to asymptotic normalization conditions
taking $\k^2 $ in the asymptotic region.  Therefore one will only obtain
information about the invariant charge in the asymptotic region where the
momenta are large compared to the masses. Because we have fixed all the
physical masses the asymptotic invariant charge of the spontaneously broken
model is not equivalent to the one of the massless symmetric model especially
as far as dependence on the masses and the normalization point is concerned.

The 1-loop $\b$-functions of the model in the asymptotic normalization are
given by $$ \eqalign{& \tilde \b _{\l,\SMALL {as}} ^{\scriptscriptstyle (1)} =
\b _ \l ^{\scriptscriptstyle (1)} = {1 \over 8 \pi ^ 2} {1 \over 3} \bigl(5 -
8 {m_{\hbox{\eightpoint $f$}} ^4 \over m_{\hbox{\eightpoint $H$}} ^4} + 4
{m_{\hbox{\eightpoint $f$}}^2 \over m _{\hbox{\eightpoint $H$}}^2} \bigr) \l^2
\equiv b _\l ^{\scriptscriptstyle (1)} ( \frac {\mf }{\mhi } ) \l^2 \cr & \b
_{ m_{\hbox{\eightpoint $f$}}} ^{\scriptscriptstyle (1)} = - {1 \over 16 \pi ^
2} { 1\over 3} \Bigl( 5 - 8 {m_{\hbox{\eightpoint $f$}} ^4\over
m_{\hbox{\eightpoint $H$}}^4} \Bigr) \l \equiv b_ {\mf}^{\scriptscriptstyle
(1)} ( \frac {\mf }{\mhi } ) \l } \EQN{\HCSP} $$ By subtracting the
RG-equation and the CS-equation one can show that the $\b$-function of the
CS-equation $ \bl $ is identical to the $\b $-function of the RG-equation
$\tilde \b _{\l,\SMALL {as}} $ for asymptotic normalization conditions.  But
in this case, where the CS- and the RG-equation have not the same structure,
they are allowed to depend on $\k $ (cf.~\quref{\KrAs}).  $$ \tilde \b
_{\l,\SMALL {as}} (\frac {\mhi ^2}{\k ^2} , \frac {\mf }{\mhi } ) = \b _
{\l,\SMALL {as}} (\frac {\mhi ^2}{\k ^2 } , \frac {\mf }{\mhi } ) \EQN\basb $$

As a simple check on the difference to the models considered in section 4, we
integrate the RG-equation with the 1-loop $\b$-function \queq{\HCSP} $$\bigl(
\kdk + b_\l ^{\scriptscriptstyle (1)} (\alpha ) \p_ \l \bigr) Q (\t , u , \a ,
\l ) = 0 \EQN{\RGSP1l} $$ with the asymptotic result $$ \widebar Q _{1,{\SMALL
{as} }} = {\l \over 1 - \frac 12 \l b_\l ^{\scriptscriptstyle (1)} (\a ) \ln
\t } = \sum _{ k=0 } ^\infty \l^ {k+1 } \bigl( \frac 12 b_\l
^{\scriptscriptstyle (1)} (\a ) \ln \t \bigr) ^k \EQN{\RGSOLSP} $$ Thereby we
have denoted $$ \t = \frac {p^2}{\k^2}\, , \quad u = \frac {\mhi ^2}{p^2 } \,
, \quad \a = \frac {\mf}{\mhi } \SUBEQNBEGIN{\DEFsp} $$ Whereas in the models
of section 4 RG-invariants are respected by the CS-equation, it is obvious
that the RG-invariant $\widebar Q_{1 , {\SMALL {as}}}$ is not a solution of
the CS-equation from 2-loop order onwards but is broken by hard terms: $$
\eqalign{ \C \widebar Q _{1, {\SMALL {as} }} = & \, \bigl( \kdk + \mhi \p
_{\mhi} + \b _{\mf } ^{\scriptscriptstyle (1)} \a \p _\a + \b _\l
^{\scriptscriptstyle (1)} \bigr) \widebar Q _{1, {\SMALL {as} }} = \b _{\mf}
^{\scriptscriptstyle (1)} \a \p _\a \widebar Q _{1, {\SMALL {as} }} \cr = & \,
\l ^3 b_{\mf } ^{\scriptscriptstyle (1)}( \a ) \a \p _ \a b _\l
^{\scriptscriptstyle (1)} (\a ) + O (\l^4 ) \neq 0 } \EQN{\CSbreak} $$

In order to get insight into the problems arising thereby we turn to the
finite RG-transformations as formulated in sect.~3.  As we have pointed out in
section 3.2, a further dependence on the mass ratio $ \a $ does not affect the
analysis of section 3.1, as long as all masses are normalized on-shell.
Therefore the structure of the 4 lowest orders is given by \queq{{\RF1l},
\shorttag {\RF2l} } and \queq{\shorttag {\RF34l} }, where now the RG-functions
depend on $\a $, too: $$ g_i (y) \longrightarrow g_i (y ,\a ) \EQN\RGFSP $$
Therefore we have $$\eqalign{ Q^{\scriptscriptstyle (1)}(\t , u , \a , \l ) &
\,= \l^2 \Bigl(g_1(u \t ,\a ) - g_1 (u , \a) \Bigr) \cr Q^{\scriptscriptstyle
(2)}(\t , u , \a , \l ) & \, = \l^3 \Bigl( g_2 (\t u, \a) - g_2(u, \a ) +
(g_1(\t u, \a ) - g_1 (u, \a ))^2 \Bigr) } \EQN{\RF1lsp} $$ and respectively
for the 3 and 4-loop order.  The $\b $-functions of the RG-equation are
therefore likewise determined by the expressions \queq{\BR1 , \RGBn }, where
the ordinary derivative is replaced by a partial one with respect to $ u \t$,
e.g.: $$\eqalign{ \tilde \b_\l ^{\scriptscriptstyle (1)} (y , \a) = & \,2
y\p_y g_1 (y ,\a ) \, \l^2 \cr \tilde \b ^{\scriptscriptstyle (2)}_\l (y , \a
) = & \, 2y \p_ y g_2 (y, \a ) \,\l^3 \cr} \EQN{\BRsp} $$ The string structure
and the $\b $-functions belonging to it are the only information contained in
finite RG-invariance. In order to get restrictions on the RG-functions $g_i
(y,\a )$ we have -- as above -- to use the CS-equation.  From now on we
restrict our considerations to an asymptotic normalization point and
consequently asymptotic momenta, which means for the 1-loop expression $$ Q ^
{\scriptscriptstyle (1) } _{\SMALL {as}} = \frac 12 \l^2 b _ \l
^{\scriptscriptstyle (1)} ( \a ) \bigl( \ln | \frac {m^2}{\k^2} |- \ln |\frac
{m^2}{p^2} | \bigr) \EQADV\Qspas1\SUBEQNBEGIN{\Qspas1a} $$ and therefore $$
g_{1,{\SMALL {as}}} (y ) = \frac 12 b _ \l ^{\scriptscriptstyle (1)} ( \a )
\ln (- y) \SUBEQN{\Qspas1b } $$ It fulfills the asymptotic CS-equation in
1-loop with the same $\b $-function as in the RG-equation (cf.~\queq\HCSP ).
We apply the CS-equation on the 2-loop RG-invariant \queq{\RF1lsp} taking into
account that the right-hand side, the soft insertion, is vanishing and find
$$\eqalign{ \b _{\l , \SMALL {as}} ^{\scriptscriptstyle (2)} (u \t , \a ) = \l
^3 & \Bigl( 2 u \p _{u } g_{2 , \SMALL {as}} (u, \a ) - \frac 12 b _{\mf} (\a)
\p_ \a b ^{\scriptscriptstyle (1)} _\l \ln (- u ) \cr & + \ln (-u \t ) \frac
12 b _{\mf} (\a) \p_ \a b ^{\scriptscriptstyle (1)} _\l (\a ) \Bigr)}
\EQN\CS2sp $$ Because the CS-equation is seen to exist from the general proof
in \quref{\CalSym } we know that the momentum dependent terms have to sum up
to a constant -- these are the same arguments as in section 4 -- , i.e.: $$
g_{2 , \SMALL {as}} (u, \a ) = - \frac 18 b ^{\scriptscriptstyle (1)}_{\mf}
(\a) \p_ \a b ^{\scriptscriptstyle (1)} _\l (\a) \ln ^2 ( -u ) + \frac 12 b
^{\scriptscriptstyle (2)} _\l (\a ) \ln ( -u) \EQN{\g2sp} $$ where we have
fixed the integration constant to zero (cf.~comments to \queq{\AB1}).
Therefore the 2-loop RG-function will start with a quadratic term in the
logarithm of the same power as $ g_{1,\SMALL {as} } ^2 (y ) $.  Summarizing
the results we find for the 2-loop invariant charge: $$\eqalign{ Q
^{\scriptscriptstyle (2)}_{\SMALL {as}} ( \frac {p^2}{\k^2} , \frac {\mhi
^2}{p^2} ,\a ,\l ) = \l^3 & \,\biggl( \frac 12 b ^{\scriptscriptstyle (2)} _\l
(\a ) \ln ( \frac {p^2}{\k^2} ) - \frac 18 b _{\mf} ^{\scriptscriptstyle (1)}
(\a) \p_ \a b ^{\scriptscriptstyle (1)} _\l (\a) \Bigl(\ln ^2 |\frac
{\mhi^2}{\k^2} | - \ln ^2 | \frac {\mhi^2}{p^2} | \Bigr) \cr & + \Bigl(b
^{\scriptscriptstyle (1)} _\l (\a) \ln ( \frac {p^2}{\k^2} ) \Bigr)^2 \biggl)
} \EQN{\Q2lsp} $$ and the $\b $-function depends on $\ln | \frac {m^2}{\k^2} |
$ $$ \b_{\l ,{\SMALL {as}} } ^{\scriptscriptstyle (2)} = \tilde \b_{\l
,{\SMALL {as}} } ^{\scriptscriptstyle (2)} = \l^3 \Bigl( - \frac 12 b _{\mf}
^{\scriptscriptstyle (1)} (\a) \a \p_ \a b ^{\scriptscriptstyle (1)} _\l (\a)
\ln | \frac {\mhi^2}{\k^2} | + b ^{\scriptscriptstyle (2)} _\l (\a) \Bigr)
\EQN\B2lsp $$ $b ^{\scriptscriptstyle (2)} _\l (\a) $ is a true 2-loop
function.  In contrast to the scalar models the 2-loop function starts with a
quadratic term in the logarithm of the same power as the 1-loop induced
RG-invariant $g_{1 ,{\SMALL {as}}} ^2 (y ) $ appearing in 2-loop order.  At
the same time the $\b $-function starts to depend logarithmically on the ratio
$ {m^2 \over \k^2 } $. Interestingly enough, an asymptotic theory in the sense
of mass-independence, if the normalization point and all momenta are taken at
infinity, does not exist: The asymptotic normalization conditions are defined
by the requirement that the terms of order ${m^2 \over \k^2 } \ln | {m^2 \over
\k^2 } | $ can be neglected. But the smaller these terms are chosen, the
larger the logarithmic term $ \ln {m^2 \over \k^2} $ in the 2-loop invariant
charge will grow.

The 2-loop RG-function we have calculated in \queq{\B2lsp} is again in
agreement with the consistency equation \queq{\CONS} tested for the invariant
charge $$ \kdk \b ^{\scriptscriptstyle (2)} _{\l,\SMALL {as}} = \b
^{\scriptscriptstyle (1)} _{\mf} \a \p _\a \b ^{\scriptscriptstyle (1)} _{\l ,
\SMALL {as}} \EQN{\Cons2ll} $$ For completeness we want to give also the
3-loop order $\b $-functions and the RG-function $ g_{3, \SMALL {as}} (y) $.
The calculation works as it did in 2-loop order, whereby now also the
$\k$-dependent part of the $\b $-function $\b ^{\scriptscriptstyle (2)} _{\mf
,\SMALL {as}} $ is determined: $$\eqaligntag{ \b ^{\scriptscriptstyle (2)}
_{\mf ,\SMALL {as}} = \, \l^2 \Bigl( & \frac 12 b _{\mf} ^{\scriptscriptstyle
(1)} (\a) b ^{\scriptscriptstyle (1)} _\l (\a) \ln | \frac {m^2}{\k^2} | + b
_{\mf} ^{\scriptscriptstyle (2)} (\a) \Bigr) &
\EQADV\Basmf3\SUBEQBEGIN{\Basmf3a} \cr \hbox{and } \qquad \qquad \quad & & \cr
\b _{\l , \SMALL {as}} ^{\scriptscriptstyle (3)} \,\,\,= \, \l^4 \Bigl( & -
\frac 1{8} b _{\mf} ^{\scriptscriptstyle (1)} (\a) b ^{\scriptscriptstyle (1)}
_\l (\a) \a \p _\a b ^{\scriptscriptstyle (1)} _\l (\a) \ln ^2 | \frac
{m^2}{\k^2} | & \SUBEQ{\Basmf3b } \cr & + \frac 18 b _{\mf}
^{\scriptscriptstyle (1)} (\a) \a \p _\a \Bigl( b _{\mf} ^{\scriptscriptstyle
(1)} (\a) \a b ^{\scriptscriptstyle (1)} _\l (\a) \Bigr) \ln ^2 | \frac
{m^2}{\k^2} | & \cr & - \frac 12 \Bigl( b _{\mf} ^{\scriptscriptstyle (1)}
(\a) \a \p_ \a b ^{\scriptscriptstyle (2)} _\l (\a) + b _{\mf}
^{\scriptscriptstyle (2)} (\a) \a \p_ \a b ^{\scriptscriptstyle (1)} _\l
(\a)\Bigr) \ln | \frac {m^2}{\k^2} | & \cr & + b_ \l ^{\scriptscriptstyle (3)}
(\a ) \Bigr) & \cr } $$ and $g_{3 ,\SMALL {as}} (y ,\a ) $ is calculated to be
$$\eqalign{ g_{3 ,\SMALL {as}} (y ,\a ) = & \, \frac 1{24 \cdot 2} b _{\mf}
^{\scriptscriptstyle (1)} (\a) \a \p _\a \Bigl( b _{\mf} ^{\scriptscriptstyle
(1)} (\a) \a b ^{\scriptscriptstyle (1)} _\l (\a) \Bigr) \ln ^3 ( -y ) \cr - &
\, \frac 1{24 \cdot 4} b _{\mf} ^{\scriptscriptstyle (1)} (\a) b
^{\scriptscriptstyle (1)} _\l (\a) \a \p _\a b ^{\scriptscriptstyle (1)} _\l
(\a) \ln ^3 ( -y )\cr - &\, \frac 18 \Bigl( b _{\mf} ^{\scriptscriptstyle (1)}
(\a) \a \p_ \a b ^{\scriptscriptstyle (2)} _\l (\a) + b _{\mf}
^{\scriptscriptstyle (2)} (\a) \a \p_ \a b ^{\scriptscriptstyle (1)} _\l
(\a)\Bigr) \ln ^2 ( -y ) \cr + & \, \frac 12 b_\l ^{\scriptscriptstyle (3)}
(\a) \ln ( -y ) } \EQN{\fastende} $$ As expected the 3-loop RG-invariant
depends on $\ln {m^2 \over \k^2} $ to the third power and the $\b $-function
to the second power.  Note that the $\b $-function $\b_ \l $ of three loop
order is not just given by the differentiation of $ g_3 (y) $ but has
according to \queq{\BRsp} an anti-symmetric contribution of the 1- and 2-loop
order.  $Q^{\scriptscriptstyle (3)}_{\SMALL {as}} $ can immediately be
calculated by inserting \queq{\fastende, \g2sp,\Qspas1 \Qspas1b } into the
3-loop expression of \queq{\RF34l}.  The $\b $-functions \queq{\Basmf3} are in
agreement with the consistency equation: $$\eqalign{ \kdk \b _{\mf,\SMALL
{as}} ^{\scriptscriptstyle (2)} \p _\a Q _{\SMALL {as}} ^{\scriptscriptstyle
(1)} = & \, - \b _\l ^{\scriptscriptstyle (1)} \p_ \l \b _{\mf}
^{\scriptscriptstyle (1)} \p _{\a} Q_{\SMALL {as}} ^{\scriptscriptstyle (1)}
\cr \kdk \b _{\l , \SMALL {as}} ^{\scriptscriptstyle (3)} = & \, \b
_{\mf,\SMALL {as}} ^{\scriptscriptstyle (2)} \a \p _\a \b _\l
^{\scriptscriptstyle (1)} + \b _{\mf } ^{\scriptscriptstyle (1)} \a \p _\a \b
_{\l,\SMALL {as}} ^{\scriptscriptstyle (2)}} \EQN{\ende} $$ Although we have
used the RG-invariance and the CS-equation only in order to compute the
invariant charge of the scalar field in a restricted range of momenta, namely
asymptotic ones, the results show that there is a far reaching difference
between the spontaneously broken model, which contains only scalar fields, and
the one, which contains fermions: In the pure scalar model considered in
section 4 strings of lower loop induced contributions are in one to one
correspondence with the RG-invariants.  This means in particular, that one
will get a sensible result, if one puts higher order RG-invariants to zero and
calculates with the lower orders in some approximation a RG-invariant result.
The most prominent example is the string of the 1-loop induced RG-invariant as
calculated from the differential equation including the 1-loop $\b $-functions
and all higher order $\b$-function taken to be zero.  In the model with
fermions as considered in this section such an separation in RG-invariants and
e.g.~the 1-loop contribution is not possible anymore: Through the CS-equation
1-loop induced contributions appear in every RG-invariant, therefore the
approximate solution of the RG-equation \queq{\RGSP1l} neglecting all higher
order $\b $-functions does not make any sense. This result is general and not
related to the asymptotic normalization condition we have chosen for
simplification.  Another important difference to the pure scalar models is the
fact, that in presence of fermions an asymptotic limit does not exist.
Considering the three lowest orders and taking into account the consistency
equation of the CS- and the RG-equation we can conclude that in the (n+1)-loop
order there appears a logarithmic term of the ratio $m^2 \over \k^2 $ to the
${\rm n^{\rm th}} $ power $$\b^{\scriptscriptstyle (n+1)} _{\l, \SMALL {as}}
\sim \ln ^n |\frac {m^2}{\k^2 } | \EQN{\BFnlc} $$ The logarithmic dependence
on the ratio of mass and normalization point can be understood from actually
having fixed two interactions at different scales: The scalar interaction by
the choice of the normalization point and the Yukawa interaction by the
Ward-identity connected with the pole of the fermion propagator, the physical
fermion mass.  To fix couplings at different scales is a very realistic
scenario thinking at the wide range of masses, which appear in the standard
model. Concluding from the calculations above, this means, that one has to
expect under such circumstances a sensitive dependence on the point where one
has normalized the couplings, i.e.~adjusted at their experimental value.

These considerations of the spontaneously broken case have to be understood as
a first look into the usefulness of structuring Green functions with the help
of RG-invariance and the CS-equation.  For all further applications it is
unavoidable to use the complete 1-loop order with all finite diagrams. Then
one can try to estimate, which terms have to be expected necessarily in the
next order.

\chap{Conclusions} In this paper our principal aim was to gain information on
the structure of the invariant charge in a 1-coupling theory from a combined
use of the CS-equation and RG-invariance.  Although they both look similar in
the infinitesimal version, in which they are derived in perturbation theory,
their meaning is completely different.  RG-invariance is -- up to field
redefinitions -- a symmetry of the Green functions, which has to be realized
in order to make the outcome of calculations independent of the arbitrary
normalization point, one has to choose to adjust the coupling to its
experimental value. The invariance under RG-transformations thus makes the
calculations universal and is one aspect, in which a renormalizable field
theory is distinguished from an effective field theory.  In contrast to this
there does not exist a compelling physical reason to require dilatational
invariance. Dilatations are broken already classically by the mass terms and
moreover in most 4-dimensional quantum field theories by hard anomalies. But
in the models we have considered in this paper the action of dilatations on
the Green functions can be expressed in form of a partial differential
equation, the CS-equation.

As usual for every symmetry in principle RG-invariance needs to be only
realized for physical observables as for example the S-matrix, but if it holds
for the Green functions up to field redefinitions, the S-matrix elements are
invariant as a consequence. In a 1-coupling theory finite RG-invariance of the
invariant charge can be deduced from the formal integration of the
RG-equation.  Therefore the invariant charge, as it is calculated in
perturbation theory, is an invariant under RG-transformations.  Because
RG-invariance is realized only to all orders of perturbation theory, every
order of the perturbative power series induces contributions to the next one
necessarily.  We have shown that RG-invariance structures the Green functions
according to their transformation properties: In particular apart from a new
RG-invariant strings of lower order RG-functions run through all orders of
perturbation theory. The RG-invariants themselves start with arbitrary
functions depending in an unique way on the renormalization point and the
mass.  Concerning the special form of the RG-functions $g_i (y) $ or their
high-energy behavior one does not get any further information from
RG-invariance in a massive theory.

As we have shown, it is the existence of the CS-equation which restricts the
high-energy behavior of the RG-functions appearing in the structure of the
invariant charge. In such models, where the CS-equation has exactly the same
form as the RG-equation, it turns out that all RG-functions tend to single
logarithms in the asymptotic limit, and as a consequence the $\b$-functions
are mass independent in this limit. Therefore the asymptotic limit goes
smoothly into the massless theory.  Apart from this high-energy restrictions
every RG-invariant is consistent with the CS-equation by itself.  This is in
marked contrast to the $U(1)$-axial model with fermions, which get their mass
via the spontaneous breaking of the symmetry. There the CS-equation contains a
$\b$-function belonging to a physical mass differential operator.  In this
case the individual RG-invariants are not by themselves solutions of the
CS-equation, but in order to satisfy the CS-equation they have to contribute
all at the same time. In every RG-invariant there appears a 1-loop induced
contribution, which means, all RG-invariants are excited by the 1-loop
contribution, none can be neglected against the other.  Moreover the
$\b$-function of the RG-equation and the CS-equation depend in the asymptotic
limit logarithmically on the ratio of mass and normalization point and the
same happens for the invariant charge from two loop order onwards.
Consequently a massless limit in the asymptotic region does not exist anymore.

The structuring according to RG-invariants combined with the respective
high-energy behavior as derived from the CS-equation is certainly a helpful
tool in any check of calculations beyond 1-loop order. As it can be seen from
the results in section 6 there appear also in such simple models as the
spontaneously broken model with fermions plenty of differently ordered 1-loop
induced terms, which can be controlled knowing the complete 1-loop order of
the respective model.  For such practical applications it seems to be of
utmost interest to extend the considerations to general Green functions
including the anomalous dimension.  \bigskip

{\it Acknowledgements} The author is grateful to C.~Becchi, J.~Gasser,
H.~Leutwyler, P.~Minkowski and K.~Sibold for useful discussions.

\refout \save{strga} \bye